\documentclass[floatfix,showpacs]{revtex4}
\usepackage{epsf}
\usepackage{epsfig, latexsym}
\usepackage{pictex}
\usepackage{amsmath}
\usepackage{amssymb}
\usepackage{amsfonts}
\usepackage[latin1]{inputenc}

\usepackage{graphicx}

\newcommand{\insertplot}[5]{\begin{figure}
 \hfill\hbox to 0.05in{\vbox to #5in{\vfill
 \inputplot{#1}{#4}{#5}}\hfill}
 \hfill\vspace{-.1in}
 \caption{#2}\label{#3}
 \end{figure}}
 \newcommand{\inputplot}[3]{
 \special{ps: plotfile #1}
}
\newcounter{fig}   

\newcommand{\vphi}{\varphi}

\renewcommand{\a}{\alpha}
\renewcommand{\b}{\beta}
\renewcommand{\c}{\gamma}
\renewcommand{\d}{\delta}
\newcommand{\f}{\phi}
\newcommand{\e}{\mu}
\newcommand{\g}{\nu}
\renewcommand{\l}{\lambda}
\renewcommand{\t}{\theta}

\begin{document}

\title{
Rotating Boson Stars and $Q$-Balls}
% \vspace{1.5truecm}
\author{Burkhard Kleihaus, Jutta Kunz}
\author{Meike List}
\altaffiliation[Present address:]{\ ZARM, Universit\"at Bremen,
Am Fallturm, D-28359 Bremen, Germany
}
\affiliation{
Institut f\"ur Physik, Universit\"at Oldenburg, Postfach 2503
D-26111 Oldenburg, Germany
}

\pacs{04.40.-b, 11.17.+d}  

\begin{abstract}
We consider axially symmetric, rotating boson stars.
Their flat space limits represent spinning $Q$-balls.
We discuss their properties
and determine their domain of existence.
$Q$-balls and boson stars are stationary solutions
and exist only in a limited frequency range.
The coupling to gravity gives rise to a spiral-like frequency dependence
of the boson stars.
We address the flat space limit and the limit of strong
gravitational coupling.
For comparison we also determine the properties of 
spherically symmetric $Q$-balls and boson stars.
\end{abstract}

\maketitle

\section{Introduction}

Non-topological solitons \cite{lee-s} or $Q$-balls \cite{coleman}
represent stationary
localized solutions in flat space possessing a finite mass.
In the simplest case they arise, when a complex scalar field
has a suitable self-interaction, mimicking 
the interaction with other fields \cite{lee-s}.
The global phase invariance of the scalar field theory
is associated with a conserved charge $Q$, 
corresponding for instance to particle number \cite{lee-s}. 
$Q$-balls thus represent non-topological solitons with charge $Q$.

Spherically symmetric $Q$-balls 
exist only in a certain frequency range, 
$\omega_{\rm min} < \omega_s < \omega_{\rm max}$,
determined by the properties of the potential
\cite{lee-s,coleman,lee-rev}.
At a critical value of the frequency,
both mass and charge of the $Q$-balls assume their minimal value,
from where they monotonically rise towards both limiting values
of the frequency.
Considering the mass of the $Q$-balls as a function of the charge,
there are thus two branches of $Q$-balls, merging and ending
at the minimal charge and mass. 
$Q$-balls are stable along the lower branch,
when their mass is smaller than the mass of $Q$ free bosons
\cite{lee-s}.

When gravity is coupled to $Q$-balls, boson stars arise 
\cite{lee-bs,lee-rev,jetzer,ms-review}.
The presence of gravity has crucial influence 
on the domain of existence of the classical solutions.
Stationary spherically symmetric boson stars
also exist only in a limited frequency range,
${\omega}_0(\kappa) < \omega_s < \omega_{\rm max}$,
where $\kappa$ denotes the strength of the gravitational coupling.
They show, however, a different type of critical behaviour.
For the smaller frequencies the boson stars
exhibit a spiral-like frequency dependence
of the charge and the mass, approaching finite limiting values
at the centers of the spirals.
When the maximal value of the frequency is approached,
the charge and the mass of the boson stars tend to zero
\cite{lee-bs,lee-rev}.

We here determine the dependence of the 
stationary spherically symmetric boson stars
on the strength of the gravitational coupling $\kappa$ and,
in particular, determine the domain of existence
of the boson stars as a function of $\kappa$,
including the limit $\kappa \rightarrow \infty$
and the flat space limit.

Recently, the existence of rotating $Q$-balls was demonstrated \cite{volkov}.
These correspond to stationary localized solutions in flat space possessing
a finite mass and a finite angular momentum. 
Interestingly, their angular momentum is quantized,
$J=nQ$ \cite{volkov,schunck}. Possessing even or odd parity,
their energy density forms one or more tori.

Rotating boson stars exhibit angular momentum quantization as well,
$J=nQ$ \cite{schunck}. 
Previously they have been obtained only
for potentials without solitonic flat space solutions,
and only for a restricted frequency range
\cite{schunck,schunck2,schunck3,japan}.
Here we consider rotating boson stars which possess flat space
counterparts. 
Our main objectives are to clarify, whether rotating boson stars exhibit
an analogous frequency dependence as non-rotating boson stars,
and to determine their domain of existence.
For that purpose, we compute numerically sequences of rotating boson stars
for constant values of the gravitational coupling strength,
focussing on fundamental solutions with
rotational quantum number $n=1$ and even parity.

In section II we recall the action, the general equations of motion
and the global charges.
In section III we present the stationary axially symmetric ansatz 
for the metric and the scalar field, we evaluate
the global charges within this Ansatz, and present the
boundary conditions for the metric and scalar field functions.
We discuss stationary spherically symmetric $Q$-ball 
and boson star solutions in section IV,
and present our results for rotating $Q$-ball and boson star solutions
in section V.
Section VI gives our conclusions.
In the Appendices we address two auxiliary functions employed in
the numerical integration (A), we present the systems of
differential equations (B) and the components of the
stress-energy tensor (C).

\section{Action, Equations and Global Charges}\label{c1}

\subsection{Action}\label{c1s1}

\begin{figure}[h!]
\centering
\includegraphics[width=70mm,angle=0,keepaspectratio]{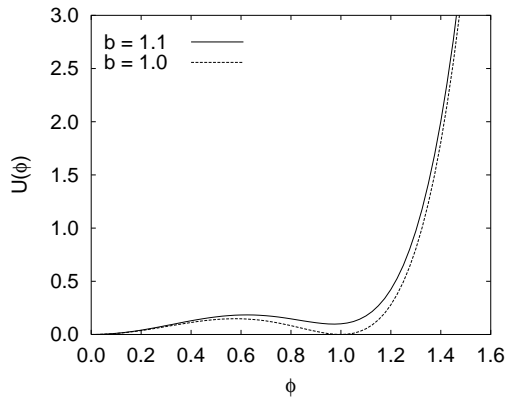}
\caption{The potential $U(\f)$ is shown
for $\lambda=1$, $a=2$ and $b=1.1$ resp.~$b=1$.}
\label{potentiale}
\end{figure}

We consider the action of a self-interacting complex scalar field 
$\Phi$ coupled to Einstein gravity
\begin{equation}
S=\int \left[ \frac{R}{16\pi G}
   -\frac{1}{2} g^{\mu\nu}\left( \Phi_{, \, \mu}^* \Phi_{, \, \nu} + \Phi _
{, \, \nu}^* \Phi _{, \, \mu} \right) - U( \left| \Phi \right|) 
 \right] \sqrt{-g} d^4x
\ , \label{action}
\end{equation}
where $R$ is the curvature scalar,
$G$ is Newton's constant,
the asterisk denotes complex conjugation,
\begin{equation}
\Phi_{,\, {\mu}}  = \frac{\partial \Phi}{ \partial x^{\mu}}
 \ ,
\end{equation}
and $U$ denotes the potential
\begin{equation}
U(|\Phi|) =  \l |\Phi|^2 \left( |\Phi|^4 -a |\Phi|^2 +b \right)
=\lambda \left( \f^6- a\f^4 + b \f^2 \right) \ ,
\label{U} \end{equation}
with $|\Phi|=\f$.
The potential is chosen such that nontopological soliton solutions
\cite{lee-s}, also referred to as $Q$-balls \cite{coleman},
exist in the absence of gravity.
As seen in Fig.~\ref{potentiale}, 
the self-interaction of the scalar field has an attractive component,
and the potential has a minimum, $U(0)=0$, at $\Phi =0$
and a second minimum at some finite value of $|\Phi|$.
The boson mass is given by $m_{\rm B}=\sqrt{\lambda b}$.

\subsection{Equations}\label{c1s2}

Variation of the action with respect to the metric
leads to the Einstein equations
\begin{equation}
G_{\mu\nu}= R_{\mu\nu}-\frac{1}{2}g_{\mu\nu}R = \kappa T_{\mu\nu}
\ , \label{Einstein}
\end{equation}
with 
$\kappa = 8\pi G$
and stress-energy tensor $T_{\mu\nu}$
\begin{eqnarray}
T_{\mu \nu} &=& \phantom{-} g_{\mu \nu} L_M 
-2 \frac{ \partial L}{\partial g^{\mu\nu}}
 \\
&=&-g_{\mu\nu} \left[ \frac{1}{2} g^{\alpha\beta} 
\left( \Phi_{, \, \alpha}^*\Phi_{, \, \beta}+
\Phi_{, \, \beta}^*\Phi_{, \, \alpha} \right)+U(\f)\right]
 + \left( \Phi_{, \, \mu}^*\Phi_{, \, \nu}
+\Phi_{, \, \nu}^*\Phi_{, \, \mu} \right
) \ .
\label{tmunu} \end{eqnarray}
Variation with respect to the scalar field
leads to the matter equation,
\begin{equation}
\left(\Box+\frac{\partial U}{\partial\left|\Phi\right|^2}\right)\Phi=0 \ ,
\label{field_phi}
\end{equation}
where $\Box$ represents the covariant d'Alembert operator. 
Equations (\ref{Einstein}) and (\ref{field_phi}) represent
the general set of Einstein--Klein--Gordon equations.

\subsection{Global Charges}\label{c1s3}

The mass $M$ and the angular momentum $J$
of stationary asymptotically flat space-times
can be obtained from their respective Komar expressions \cite{wald},
\begin{equation}
{M} = 
 \frac{1}{{4\pi G}} \int_{\Sigma}
 R_{\mu\nu}n^\mu\xi^\nu dV
\ , \label{komarM1}
\end{equation}
and
\begin{equation}
{\cal J} =  -
 \frac{1}{{8\pi G}} \int_{\Sigma}
 R_{\mu\nu}n^\mu\eta^\nu dV
\ . \label{komarJ1}
\end{equation}
Here $\Sigma$ denotes an asymptotically flat spacelike hypersurface,
$n^\mu$ is normal to $\Sigma$ with $n_\mu n^\mu = -1$,
$dV$ is the natural volume element on $\Sigma$,
$\xi$ denotes an asymptotically timelike Killing vector field
and $\eta$ an asymptotically spacelike Killing vector field
\cite{wald}.
Replacing the Ricci tensor via the Einstein equations by the
stress-energy tensor yields 
\begin{equation}
M
= \, 2 \int_{\Sigma} \left(  T_{\mu \nu} 
-\frac{1}{2} \, g_{\mu\nu} \, T_{\gamma}^{\ \gamma}
 \right) n^{\mu }\xi^{\nu} dV \ ,
 \label{komarM2}
\end{equation}
and
\begin{equation}
{\cal J} = -
 \int_{\Sigma} \left(  T_{\mu \nu}
-\frac{1}{2} \, g_{\mu\nu} \, T_{\gamma}^{\ \gamma}
 \right) n^{\mu }\eta^{\nu} dV \ .
 \label{komarJ2}
\end{equation}

A conserved charge $Q$ is associated
with the complex scalar field $\Phi$,
since the Lagrange density is invariant under the global phase transformation
\begin{equation}
\displaystyle
\Phi \rightarrow \Phi e^{i\alpha} \ 
\end{equation}
leading to the conserved current
\begin{eqnarray}
j^{\mu} & = &  - i \left( \Phi^* \partial^{\mu} \Phi 
 - \Phi \partial^{\mu}\Phi ^* \right) \ , \ \ \
j^{\mu} _{\ ; \, \mu}  =  0 \ .
\end{eqnarray}

\section{Ansatz and Boundary Conditions}\label{c2}

\subsection{Ansatz}\label{c2s1}

To obtain stationary axially symmetric solutions,
we impose on the space-time the presence of
two commuting Killing vector fields,
$\xi$ and $\eta$, where
\begin{equation}
\xi=\partial_t \ , \ \ \ \eta=\partial_{\varphi}
\   \label{xieta} \end{equation}
in a system of adapted coordinates $\{t, r, \theta, \varphi\}$.
In these coordinates the metric is independent of $t$ and $\varphi$,
and can be expressed in isotropic coordinates
in the Lewis--Papapetrou form \cite{kk}
\begin{eqnarray}
ds^2 &=&- f dt^2 
+ \frac{l}{f} \, \biggl[ g \left( dr^2 + r^2 \, d\t^2 \right) 
  + r^2 \, \sin^2 \t \,  \, \left( d \varphi
- \frac{\omega}{r} \, dt \right)^2 \biggr] \ . \label{ansatzg}
\end{eqnarray}
The four metric functions $f$, $l$, $g$ and $\omega$
are functions of the variables $r$ and $\theta$ only.

The symmetry axis of the spacetime, where $\eta=0$, 
corresponds to the $z$-axis.
The elementary flatness condition \cite{book}
\begin{equation}
\frac{X,_\mu X^{, \, \mu}}{4X} = 1 \ , \ \ \
X=\eta^\mu \eta_\mu \
\    \label{regcond} \end{equation}
then imposes on the symmetry axis the condition \cite{kk}
\begin{equation}
g|_{\theta=0}=g|_{\theta=\pi}=1 \ .
\end{equation}

For the scalar field $\Phi$ we adopt the stationary ansatz \cite{schunck}
\begin{eqnarray}
\Phi (t,r,\t, \varphi)& = & \f (r, \t)
 e^{ i\omega_s t +i n \varphi} \ , \label{ansatzp}
\end{eqnarray}
where $\f (r, \t)$ is a real function,
and $\omega_s$ and $n$ are real constants.
Single-valuedness of the scalar field requires
\begin{equation}
\Phi(\varphi)=\Phi(2\pi + \varphi) \ , 
\end{equation}
thus the constant $n$ must be an integer,
i.e., $n \, = \, 0, \, \pm 1, \, \pm 2, \, \dots$.
We refer to $n$ as rotational quantum number.
When $n \not=0$, the phase factor $\exp{(i n \varphi)}$ 
prevents spherical symmetry of the scalar field $\Phi$. 

Thus to obtain stationary axially symmetric boson stars
a system of five coupled partial differential equations
needs to be solved. This set of equations is presented in Appendix 
\ref{dgl_sys2}.
In contrast, for the $Q$-balls of flat space
the metric is the Minkowski metric, i.e.,$f=l=g=1$, $\omega=0$.
Here, at least in principle,
only a single partial differential equation for the scalar field
function needs to be solved.

For stationary spherically symmetric boson stars $n=0$,
and the scalar field function $\f$ depends only on
the radial coordinate, $\f=\f (r)$.
The metric then simplifies as well, since $g\equiv 1$ and $\omega\equiv 0$,
and the non-trivial metric functions $f$ and $l$ depend only on
the radial coordinate, $f=f(r)$, $l=l(r)$.
Thus for stationary spherically symmetric solutions
one obtains a much simpler system of three
coupled ordinary differential equations.
This set of equations is presented in Appendix 
\ref{dgl_sys1}.

\subsection{Mass, angular momentum and charge}\label{c2s2}

The mass $M$ and the angular momentum $J$ 
can be read off the asymptotic expansion of the metric functions $f$ 
and $\omega$, respectively,
\cite{kkrot1}
\begin{equation}
f = 1- \frac{2MG}{r} + O\left( \frac{1}{r^2} \right) \ , \ \ \
 \omega = \frac{2JG}{r^2} + O\left( \frac{1}{r^3} \right)\ ,
\label{MQasym1} \end{equation}
i.e.,
\begin{eqnarray}
M=\frac{1}{2G} \lim_{r \rightarrow \infty} r^2\partial_r \, f 
\ , \ \ \  J=\frac{1}{2G} \lim _{r \rightarrow \infty} r^2\omega \ .
\label{MJ2}
\end{eqnarray}
This is seen by considering the Komar expressions
Eqs.~(\ref{komarM1}) and (\ref{komarJ1}),
with unit vector $n^\mu = (1, 0, 0, \omega/r)/\sqrt{f}$,
and volume element 
$dV =1/ \sqrt{f} \, |g|^{1/2} \, dr \, d\t \, d\varphi$,
leading to \cite{kkrot2}
\begin{eqnarray}
 M &=&
 -\frac{1}{ 8 \pi G} \int_\Sigma  R_t^t \sqrt{-g} dr d\theta d\vphi 
\nonumber \\
 &=&  \lim_{r\to\infty}
 \frac{2\pi }{8\pi G} \int_0^{\pi}
 \left. \left[\frac{\sqrt{l}}{f}  r^2  \sin\theta
 \left( \frac{\partial f}{\partial r} -
 \frac{l}{f} \sin^2\theta \omega
 \left(\frac{\partial \omega}{\partial r} - \frac{\omega}{r}   \right)
 \right)
 \right]  \right|_{r} d\theta
 \ , \phantom{\frac{2\pi }{4\pi G}}
\end{eqnarray}
and similarly
\begin{equation}
J=\lim_{r\to\infty} 
\frac{2\pi }{16\pi G} \int_0^{\pi}
\left. \left[
\frac{l^{3/2}}{f^2} r^2 \sin^3\theta
\left( \omega - r \frac{\partial\omega}{\partial r} \right) 
\right]  \right|_{r} d\theta \ .
\label{localJ} \end{equation}
Insertion of the asymptotic expansions of the metric functions
then yields expressions (\ref{MJ2}). 

Alternatively, the mass $M$ and the angular momentum $J$
can be obtained by direct integration of the expressions
(\ref{komarM2}) and (\ref{komarJ2}), 
\begin{eqnarray}
M
&=& \phantom{2}\int_{\Sigma} \left( 2 T_{\e}^{\g} 
 - \d _{\e}^{\g} \, T_{\c}^{\c} \right) \, n_{\g} \, \xi^{\e} dV \ , 
\nonumber\\
&=&\int \left(2 \, T_t^t -T_{\e}^{\e} \right) \, |g|^{1/2} 
\, dr \, d\t \, d \varphi \ , \label{tolman}
\end{eqnarray}
corresponding to the Tolman mass, and
\begin{eqnarray}
J&=& -
\int T_{\, \varphi}^{\, t} \, |g|^{1/2} \, dr \, d\t \, d \varphi \ . 
\label{ang2}
\end{eqnarray}

The conserved scalar charge $Q$ is obtained from the time-component
of the current,
\begin{eqnarray}
Q &=- & \int j^t \left| g \right|^{1/2} dr d\t d\varphi 
\nonumber \\
%& = & -i\int \left( \Phi^* \dot{\Phi} - \Phi \dot{\Phi}^* \right) 
%\left|g \right|^{1/2} dr d\t d\varphi  \ ,
%\label{ladung} \end{eqnarray}
%thus $Q$ vanishes, unless $\Phi$ is time-dependent.
%Evaluation yields for the scalar charge
%\begin{eqnarray}
%Q&=& 4 \pi \omega_s \int_0^{\infty}\int _0^{\pi} 
 &=& 4 \pi \omega_s \int_0^{\infty}\int _0^{\pi} 
 |g| ^{1/2}   \frac{1}{f}  \left(  1 +
  \frac{n}{\omega_s}\frac{\omega}{r} \right) \f^2 \,
dr \, d\t 
\ . \label{Qc}
\end{eqnarray}

From Eq.~(\ref{ang2}) for the angular momentum $J$
and Eq.~(\ref{Qc}) for the scalar charge $Q$,
one obtains the important quantization relation for the angular momentum,
\begin{equation}
J=n \, Q \ , \label{JnQ}
\end{equation}
derived first by Schunck and Mielke \cite{schunck},
by taking into account that 
$T_{\, \varphi}^{\, t} = n j^t$, since
$\partial_\varphi \Phi =  i n \Phi$.
Thus a spherically symmetric boson star
has angular momentum $J =0$, because $n=0$.

\subsection{Boundary conditions}\label{c2s3}

The choice of appropriate boundary conditions must guarantee 
that the boson star solutions
are globally regular and asymptotically flat,
and that they possesses finite energy and finite energy density.

For spherically symmetric boson stars
boundary conditions must be specified
for the metric functions $f(r)$ and $l(r)$ 
and the scalar field function $\f(r)$ 
at the origin and at infinity.
At the origin one finds the boundary conditions
\begin{equation}
\partial_r f|_{r=0}=0 \ , \ \ \  
\partial_r l|_{r=0}=0 \ , \ \ \
\partial_r \f|_{r=0}=0 \ .
\label{bc1} \end{equation}
Note, that for spherically symmetric boson stars the scalar field
has a finite value $\f_0$ at the origin,
\begin{equation}
\f(r)= \f_0 \, + O(r^2 ) \ . \label{phi_r0}
\end{equation}
For $r \rightarrow \infty$ 
the metric approaches the Minkowski metric $\eta_{\a\b}$
and the scalar field assumes its vacuum value $\Phi=0$. 
Accordingly, we impose at infinity the boundary conditions 
\begin{equation}
f|_{r \rightarrow \infty}=1 \ , \ \ \
l|_{r \rightarrow \infty}=1 \ , \ \ \
\f|_{\, r \rightarrow \infty}=0 \ . 
\label{bc2} \end{equation}

For rotating axially symmetric boson stars
appropriate boundary conditions must be specified
for the metric functions $f(r,\theta)$, $l(r,\theta)$, $g(r,\theta)$
and $\omega(r,\theta)$, 
and the scalar field function $\f(r,\theta)$
at the origin, at infinity, on the positive $z$-axis ($\theta=0$),
and, exploiting the reflection symmetry w.r.t.~$\theta \rightarrow
\pi - \theta$, in the $xy$-plane ($\theta=\pi/2$).
At the origin we require
\begin{equation}
\partial_r f|_{r=0}=0 \ , \ \ \
\partial_r l|_{r=0}=0 \ , \ \ \
g|_{r=0}=1 \ , \ \ \
\omega|_{r=0}=0 \ , \ \ \
\f| _{r =0}=0 \ .
\label{bc3} \end{equation}
At infinity the boundary conditions are
\begin{equation}
f|_{r \rightarrow \infty} =1 \ , \ \ \
l|_{r \rightarrow \infty} =1 \ , \ \ \
g|_{r \rightarrow \infty} =1 \ , \ \ \
\omega|_{r \rightarrow \infty} =0 \ , \ \ \
\f| _{r \rightarrow \infty}=0 \ ,
\label{bc4} \end{equation}
and for $\t=0$ and $\t=\pi/2$, respectively, 
we require the boundary conditions
\begin{equation}
\partial_{\t} f|_{\t=0}=0 \ , \ \ \
\partial_{\t} l|_{\t=0}=0 \ , \ \ \
g|_{\t=0}=1 \ , \ \ \
\partial_{\t} \omega |_{\t=0}=0 \ , \ \ \
\f |_{\t=0}=0 \ , 
\label{bc5} \end{equation}
and for even parity solutions
\begin{equation}
\partial_{\t} f|_{\t=\pi/2}=0 \ , \ \ \
\partial_{\t} l|_{\t=\pi/2}=0 \ , \ \ \
\partial_{\t} g|_{\t=\pi/2}=0 \ , \ \ \
\partial_{\t} \omega |_{\t=\pi/2}=0 \ , \ \ \
\partial_{\t} \f |_{\t=\pi/2}=0 \ ,
\label{bc6} \end{equation}
while for odd parity solutions
$\f |_{\t=\pi/2}=0$.

\section{Stationary Spherically Symmetric Solutions}\label{c3}

Stationary spherically symmetric solutions are obtained, when
$n=0$.
The set of coupled non-linear ordinary differential equations,
given in Appendix \ref{dgl_sys1},
is solved numerically \cite{COLSYS},
subject to the above boundary conditions, Eqs.~(\ref{bc1})-(\ref{bc2}).
Because of the power law fall-off of the metric functions,
we compactify space by introducing the compactified radial coordinate
\begin{equation}
\bar r = \frac{r}{1+r} \ .
\label{rcomp} \end{equation}
The numerical calculations employ
a collocation method for boundary-value ordinary differential equations,
using the damped Newton scheme for a sequence of meshes,
until the required accuracy is reached \cite{COLSYS}.

\subsection{Solutions in flat space: $Q$-balls}

Spherically symmetric $Q$-balls have been studied before
(see e.g.~\cite{lee-s,coleman,volkov}, or \cite{lee-rev} for a review).
We here review their main features, to be able to better
demonstrate the effects of gravity and rotation.

Recalling first the mass $M$ of the $Q$-balls,
\begin{eqnarray}
\label{radenergy}
M 
& = & 4 \pi \int_0^\infty T_{tt} r^2 \, d r \ 
=
4 \pi \int_0^\infty  \left[\omega_s^2 \, \phi^2 +
\phi^{\, \prime \, 2} + U(\phi)\right] r^2 \, d r \ ,
\end{eqnarray}
where the prime denotes differentiation with respect to $r$, and
and their charge $Q$,
\begin{eqnarray}
\label{charge}
Q(\omega_s) & = & 8 \pi \, \omega_s \int_0^\infty \phi^2 \, r^2 \, d r \ ,
\end{eqnarray}
we follow the discussion in \cite{volkov} to obtain the limits
for the frequency $\omega_s$
\begin{equation}
\omega_{\rm min}^2 < \omega_s^2 < \omega_{\rm max}^2 \ ,
\label{olimits}
\end{equation}
while correcting for the missing factor of two 
in the field equation \cite{foot1,foot2}.
The equation of motion for the scalar field \cite{foot1}, 
\begin{eqnarray}
\label{radeom}
0 & = & \phi'' + \frac{2} {r} \, \phi' - 
\frac{1}{2} \frac{d U(\phi)} {d\phi} +
\omega_s^2 \, \phi \ ,
\end{eqnarray}
is equivalent to \cite{foot1}
\begin{eqnarray}
\label{radeom2}
\frac{1}{2} \, \phi^{\prime 2} + 
\frac{1}{2} \,\omega_s^2 \, \phi^2 - \frac{1}{2} U & = & {\cal E} - 2 \int_0^r
\frac{\phi^{\, \prime\, 2}}{r} \, dr \ ,
\end{eqnarray}
where ${\cal E}$ is an integration constant,
and effectively describes a particle moving with friction in the potential
\cite{foot1}
\begin{eqnarray}
V(\phi) = \frac{1}{2} \, \omega_s^2 \, \phi^2
 - \frac{1}{2} U(\phi) \ .
\label{Vpot}
\end{eqnarray}
The first necessary condition for $Q$-balls to exist
is then $V^{''}(0) < 0$ \cite{volkov}
and yields the maximal frequency $\omega_{\rm max}$ \cite{foot1}
\begin{eqnarray}
\label{cond1}
\omega_s^2 & < & \omega^{2}_{\rm max} \; \equiv \; 
\frac{1}{2} U''(0) = \lambda \, b \ ,
\end{eqnarray}
while the second condition is that $V(\f)$ should
become positive for some non-zero value of
$\f$ \cite{volkov}, i.e., \cite{foot1}
\begin{eqnarray}
\label{cond2}
\omega_s^2 & > & \omega^{2}_{\rm min} \; \equiv \; 
\min_{\f} \left[{U(\phi)}/{\phi^2} \right] \; = \; 
\lambda \left(b- \frac{a^2}{4} \right) \ .
\end{eqnarray}

Turning now to the $Q$-ball solutions, we specify the potential parameters as
\cite{volkov}
\begin{equation}
\lambda = 1 \ , \ \ \ a=2 \ , \ \ \ b=1.1 \ .
\label{param}
\end{equation}
Fixing the value of $\omega_s$ in the allowed range, one obtains a sequence
of $Q$-ball solutions, consisting of the fundamental $Q$-ball and its
radial excitations \cite{volkov}.
The boson function $\f$ of the fundamental $Q$-ball has no nodes,
while it has $k$ nodes for the $k$-th radial excitation.

\begin{figure}[h!]
\parbox{\textwidth}
{\centerline{
\mbox{
\epsfysize=10.0cm
\includegraphics[width=70mm,angle=0,keepaspectratio]{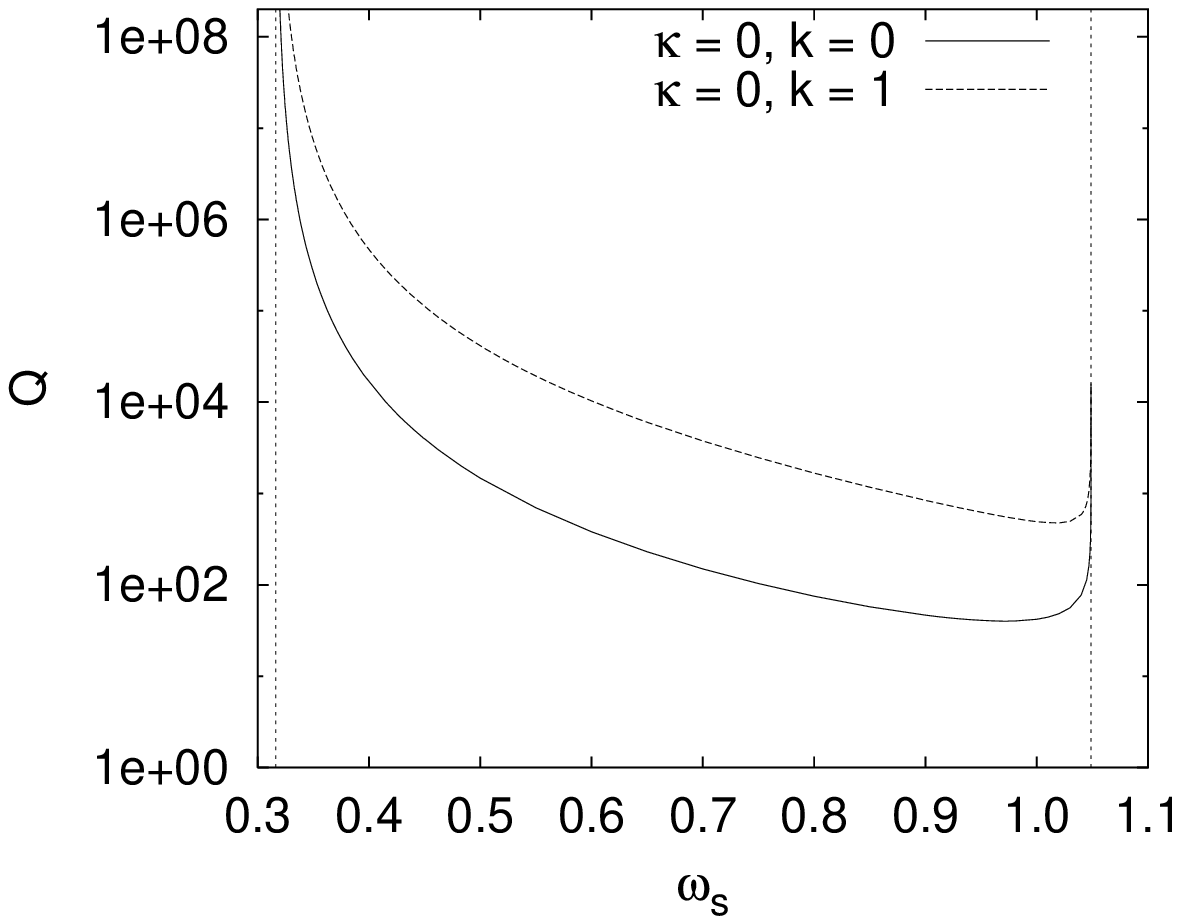}
\includegraphics[width=70mm,angle=0,keepaspectratio]{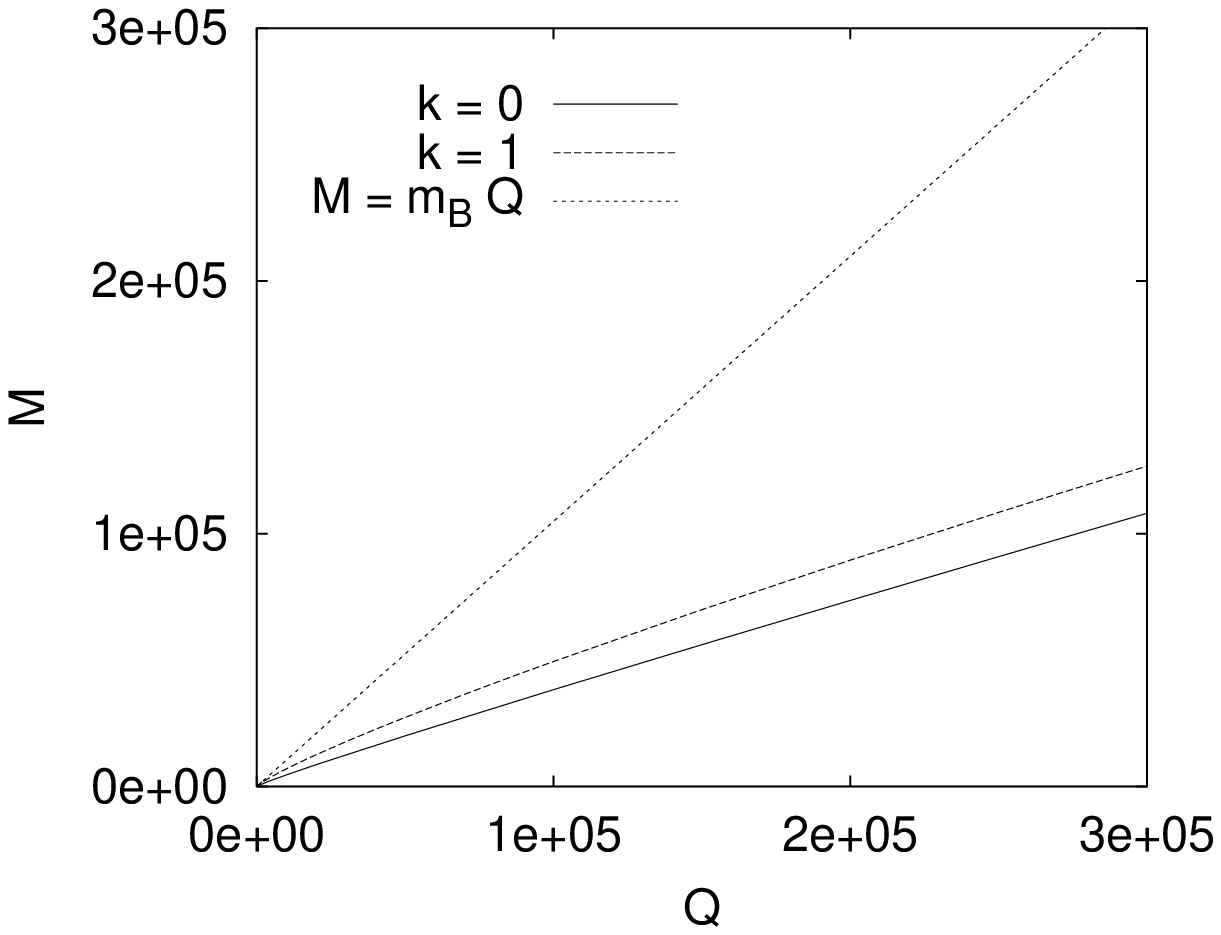}
}}}
\caption{
%2
Left: The charge $Q$ is shown as a function of the frequency $\omega_s$
for fundamental $Q$-balls ($k=0$) and their first radial excitations ($k=1$).
Also shown are the limiting values of the frequency, $\omega_{\rm min}$
and $\omega_{\rm max}$.
Right: The mass $M$ is shown as a function of the charge $Q$
for fundamental $Q$-balls ($k=0$) and their first radial excitations ($k=1$).
Also shown is the mass for $Q$ free bosons, $M=m_{\rm B}Q$.
The upper branches of the mass $M$ are not discernible (on this scale)
from the mass of $Q$ free bosons.
}
\label{Qvsomega0}
\end{figure}

Focussing on the fundamental $Q$-ball solutions and their first
radial excitations,
we exhibit in Fig.~\ref{Qvsomega0} the dependence of the charge $Q$
on the frequency $\omega_s$.
As seen in the figure, at a critical value
$\omega_{\rm cr}$ the charge assumes its minimal value $Q_{\rm cr}$.
The charge diverges both when $\omega_s \rightarrow \omega_{\rm min}$
and when $\omega_s \rightarrow \omega_{\rm max}$ \cite{lee-s}.
%\cite{lee-s,foot3}.

\begin{figure}[h!]
\parbox{\textwidth}
{\centerline{
\mbox{
\epsfysize=10.0cm
\includegraphics[width=70mm,angle=0,keepaspectratio]{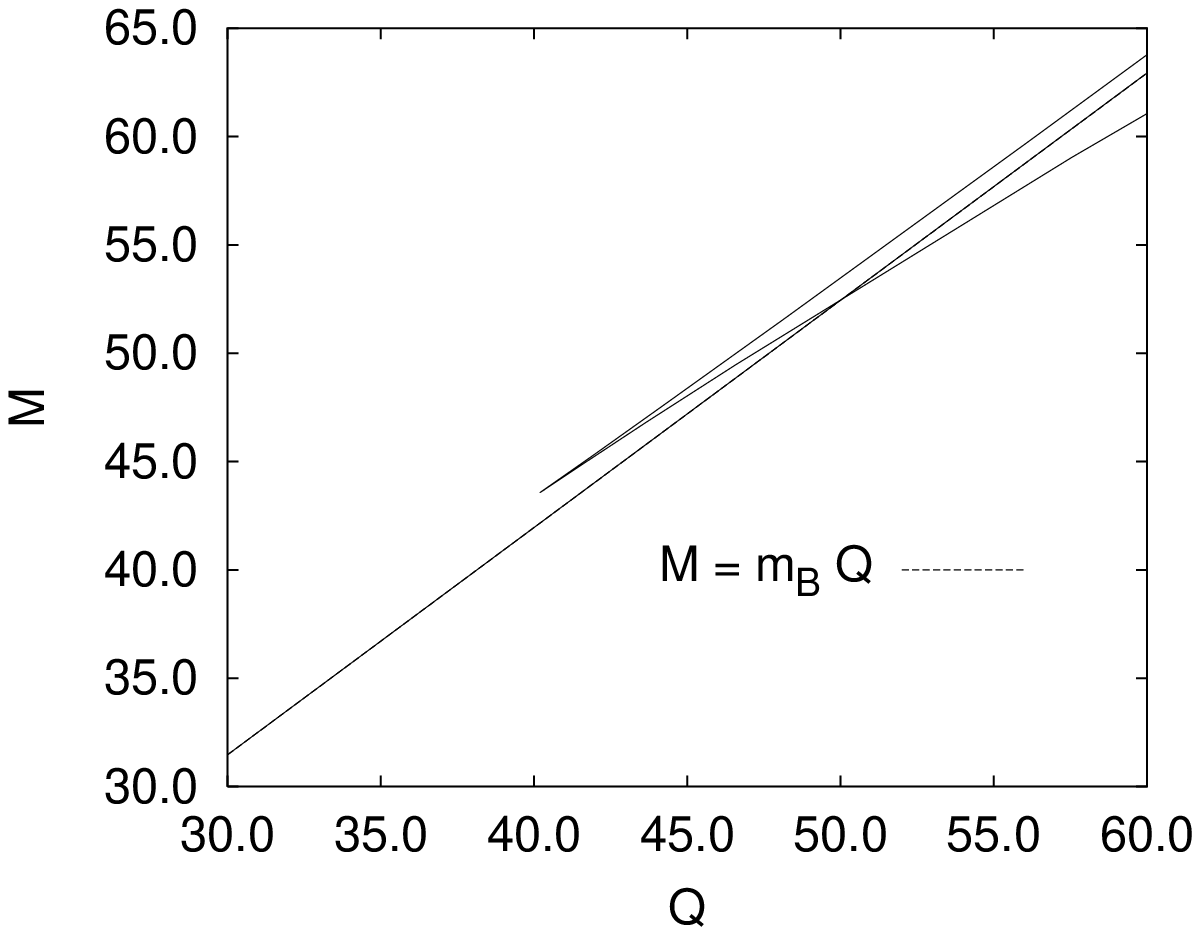}
\includegraphics[width=70mm,angle=0,keepaspectratio]{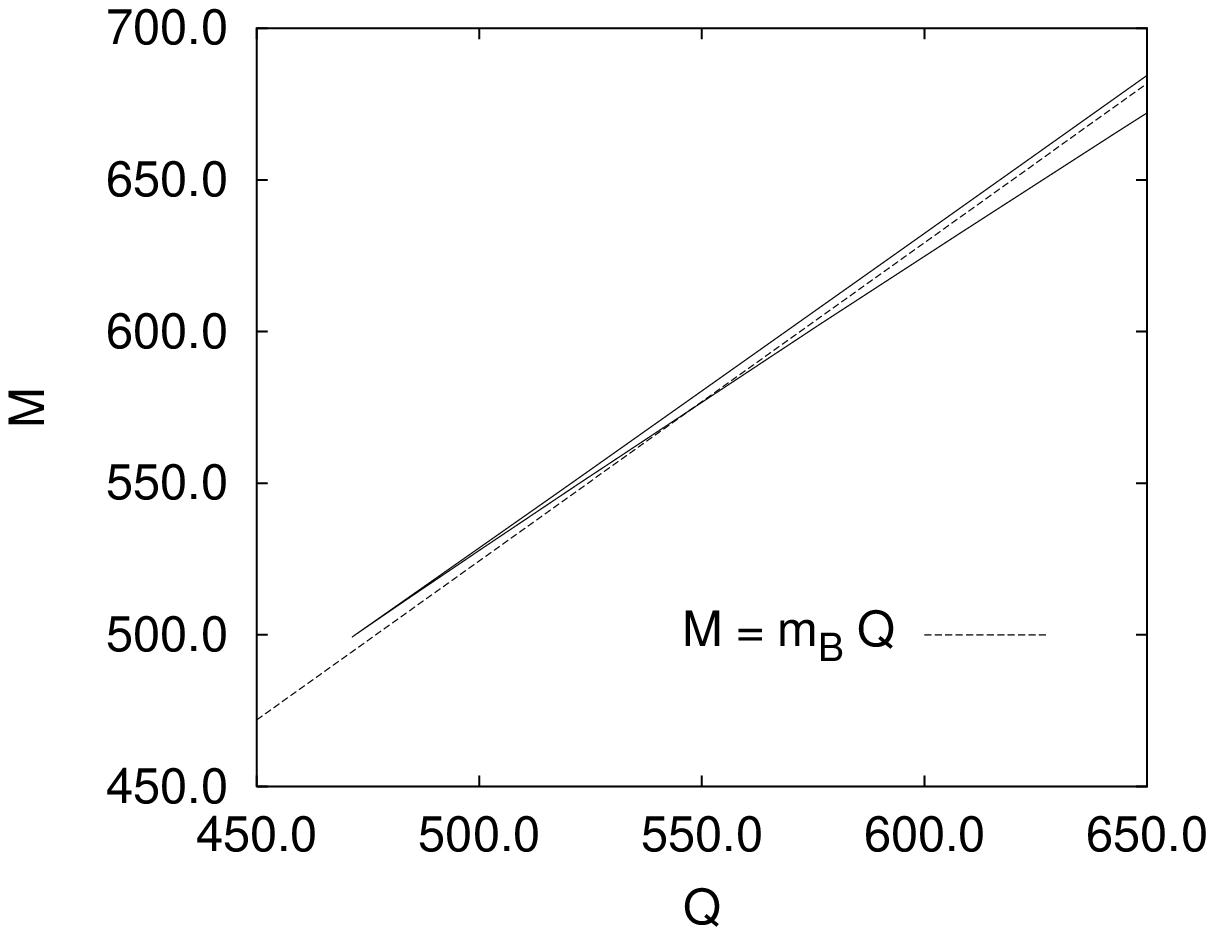}
}}}
\caption{
%3
The mass $M$ is shown as a function of the charge $Q$
for fundamental $Q$-balls ($k=0$, left) 
and their first radial excitations ($k=1$, right)
in the region close to their corresponding critical values $Q_{\rm cr}$.
Also shown is the mass for $Q$ free bosons, $M=m_{\rm B}Q$.
}
\label{MvsQa}
\end{figure}

Fig.~\ref{Qvsomega0} also exhibits the mass $M$ 
as a function of the charge $Q$.
The region close to the critical value $Q_{\rm cr}$ 
is exhibited in Fig.~\ref{MvsQa} 
for the fundamental $Q$-balls and their first radial excitations.
The lower branches correspond to values of the frequency
$\omega_s < \omega_{\rm cr}$, while the upper branches
represent the values $\omega_s > \omega_{\rm cr}$.
When the mass is smaller than the mass of $Q$ free bosons,
$M<m_{\rm B}Q$, the solutions are stable \cite{lee-s}.
When $\omega_s \rightarrow \omega_{\rm max}$,
the upper branches approach the mass $M=m_{\rm B}Q$ of $Q$ free bosons
from above \cite{lee-s}.

\begin{figure}[h!]
\parbox{\textwidth}
{\centerline{
\mbox{
\epsfysize=10.0cm
\includegraphics[width=70mm,angle=0,keepaspectratio]{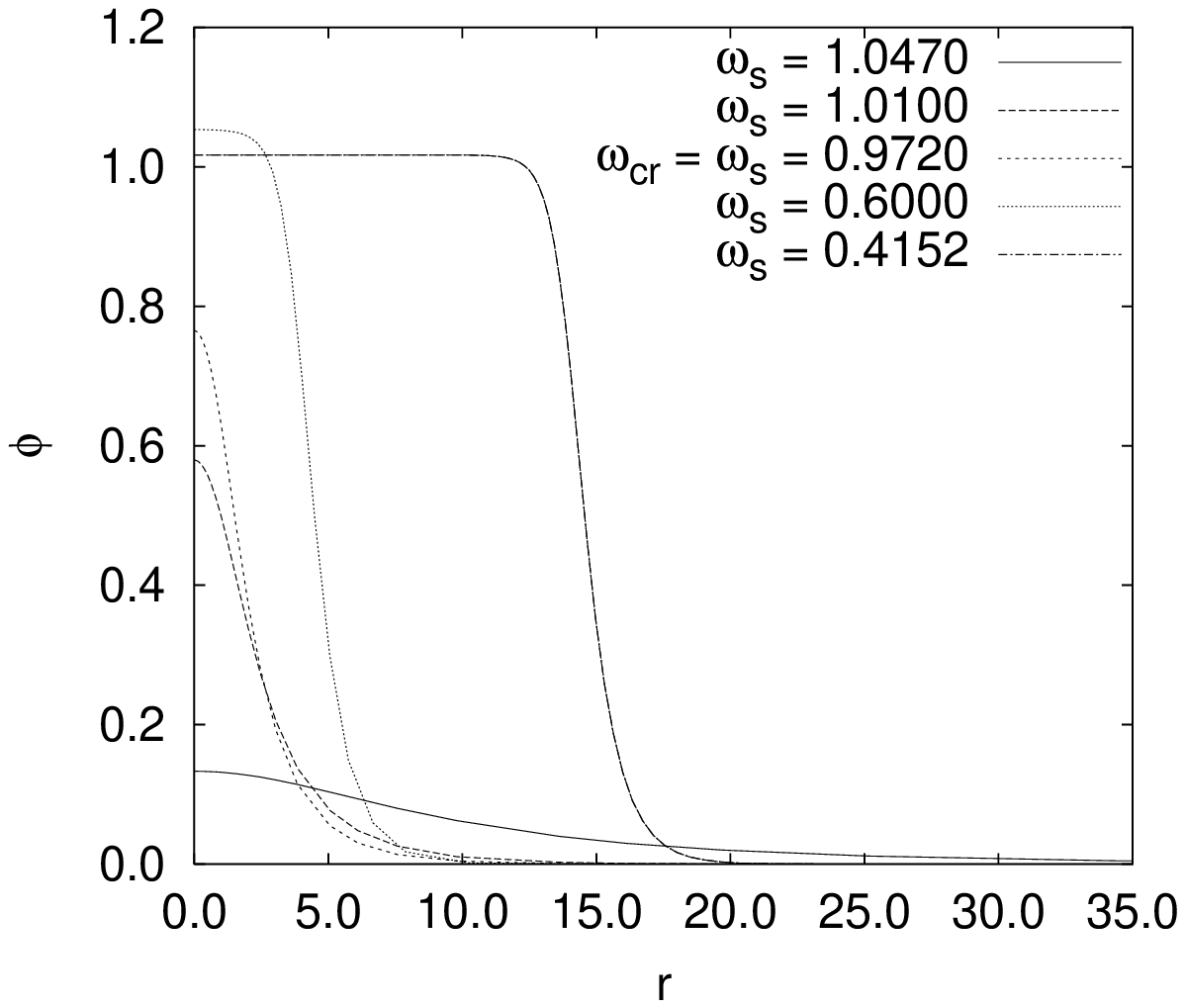}
\includegraphics[width=70mm,angle=0,keepaspectratio]{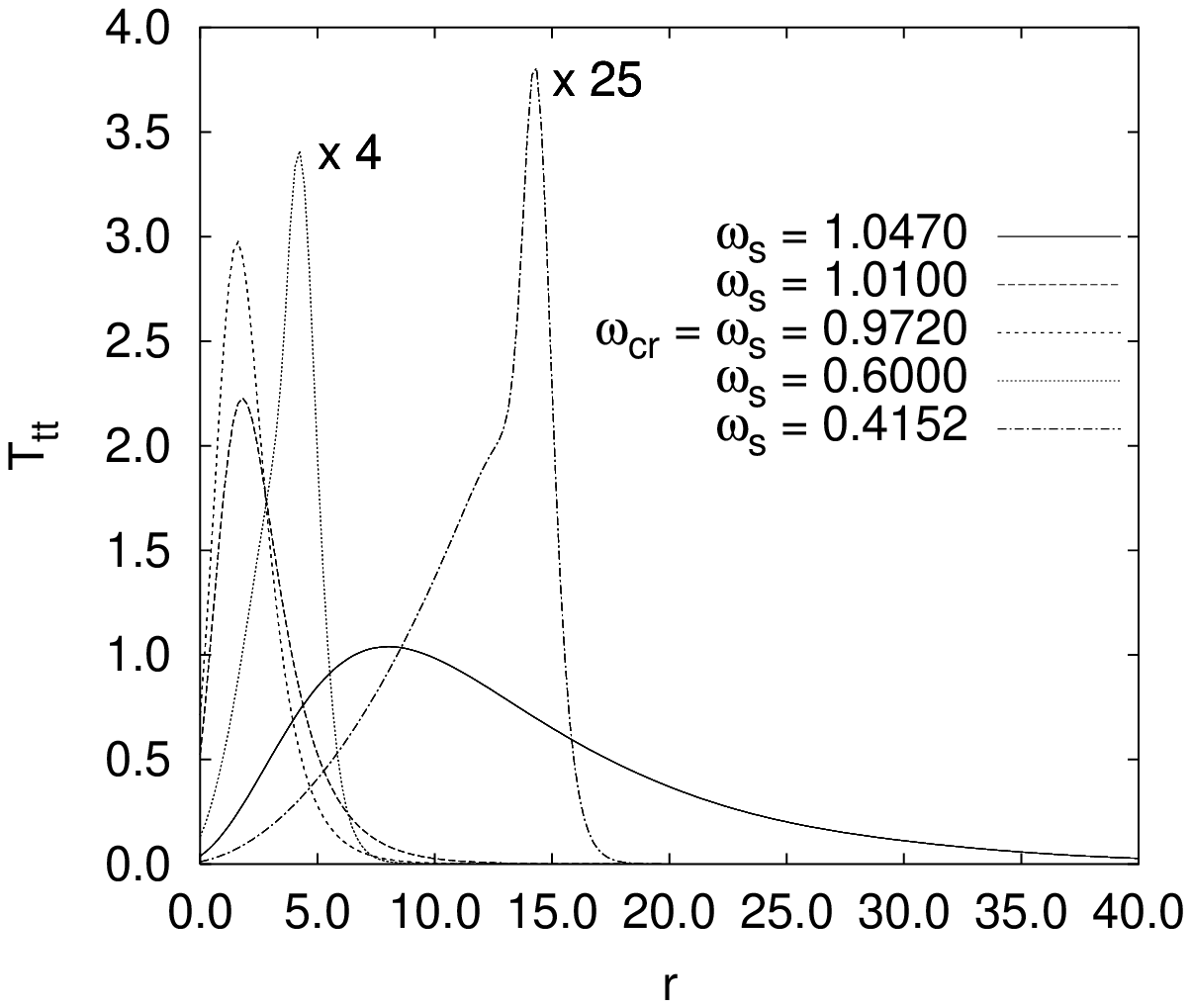}
}}}
\caption{
%4
The scalar field $\f(r)$ and the energy density $T_{tt}(r)$ 
are shown for fundamental $Q$-balls ($k=0$)
for several values of the frequency $\omega_s$ including $\omega_{\rm cr}$.
For $\omega_s=0.6$ and $\omega_s=0.4152$ the scaled
energy density $T_{tt}(r)/4$ resp.~$T_{tt}(r)/25$ is shown.
}
\label{ohnerot_1}
\end{figure}

The scalar field function $\f$ and the energy density $T_{tt}$
of the fundamental $Q$-balls
are shown in Fig.~\ref{ohnerot_1} for 
several values of the frequency $\omega_s$ including $\omega_{\rm cr}$.
The energy density of the $Q$-balls is shell-like.
Along the upper branch, with increasing charge and mass
the maximum of the energy density decreases while moving outwards,
whereas along the lower branch, with increasing charge and mass
the maximum of the energy density increases while moving outwards,
being strongly correlated with the steep fall-off
of the scalar field function.
Radially excited $Q$-balls with $k$ nodes
possess $k+1$ energy shells \cite{volkov}.

\subsection{Solutions in curved space: boson stars}

When the scalar field is coupled to gravity, boson stars arise
(see e.g.~\cite{jetzer,schunck2} for reviews).
Spherically symmetric boson stars, based on a self-interacting
boson field with a potential $U$, Eq.~(\ref{U}),
have been considered by Friedberg, Lee, and Pang \cite{lee-bs}. 
For their choice of the potential parameters 
the minima of the potential are degenerate,
yielding for the minimum value of the frequency $\omega_{\rm min}=0$.
In contrast, for our choice of parameters, Eq.~(\ref{param}),
the potential has a global and a local minimum,
and $\omega_{\rm min}>0$.
The different form of the potential has consequences for some 
features of the boson stars, as discussed below.

\begin{figure}[h!]
\parbox{\textwidth}
{\centerline{
\mbox{
\epsfysize=10.0cm
\includegraphics[width=70mm,angle=0,keepaspectratio]{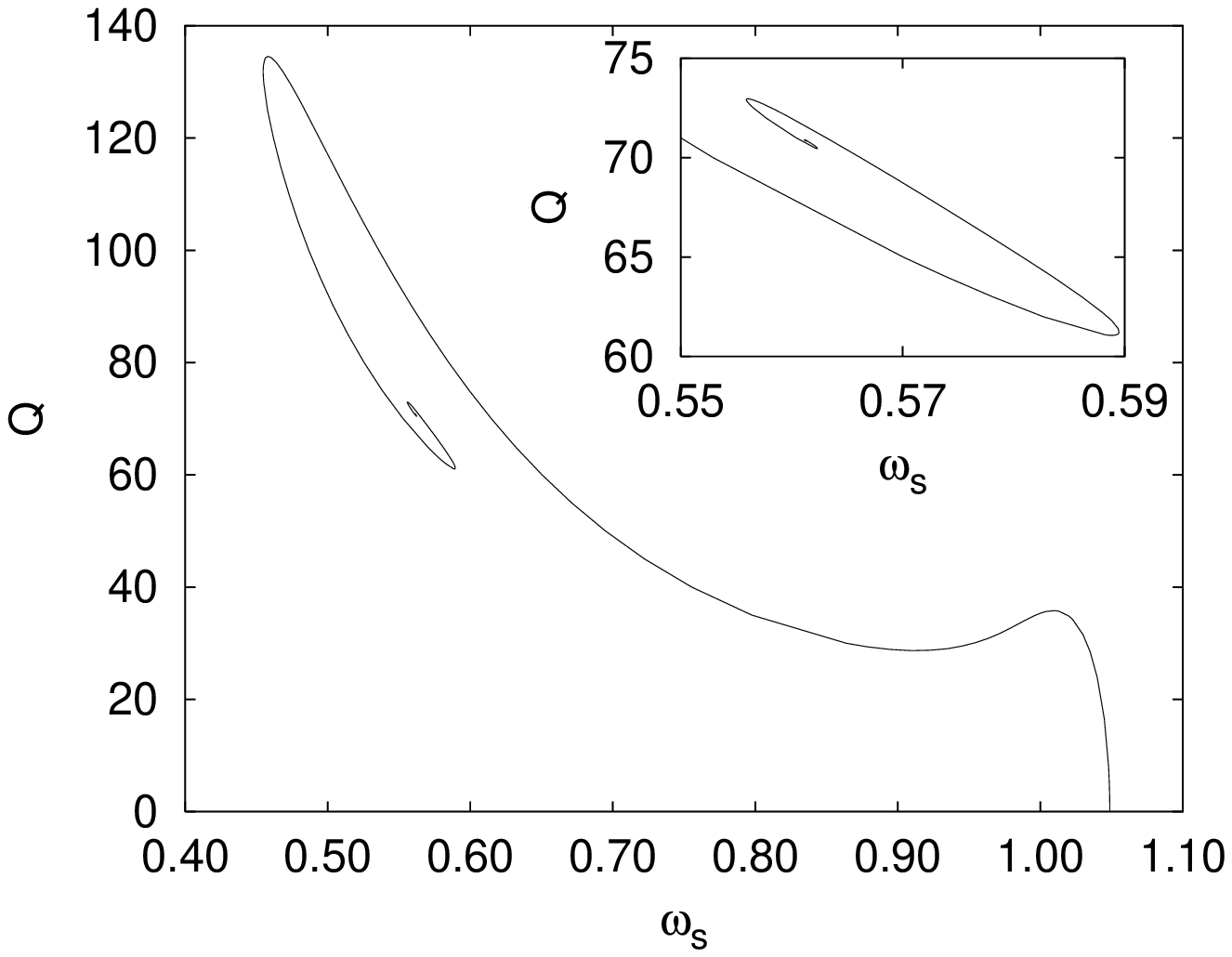}
\includegraphics[width=70mm,angle=0,keepaspectratio]{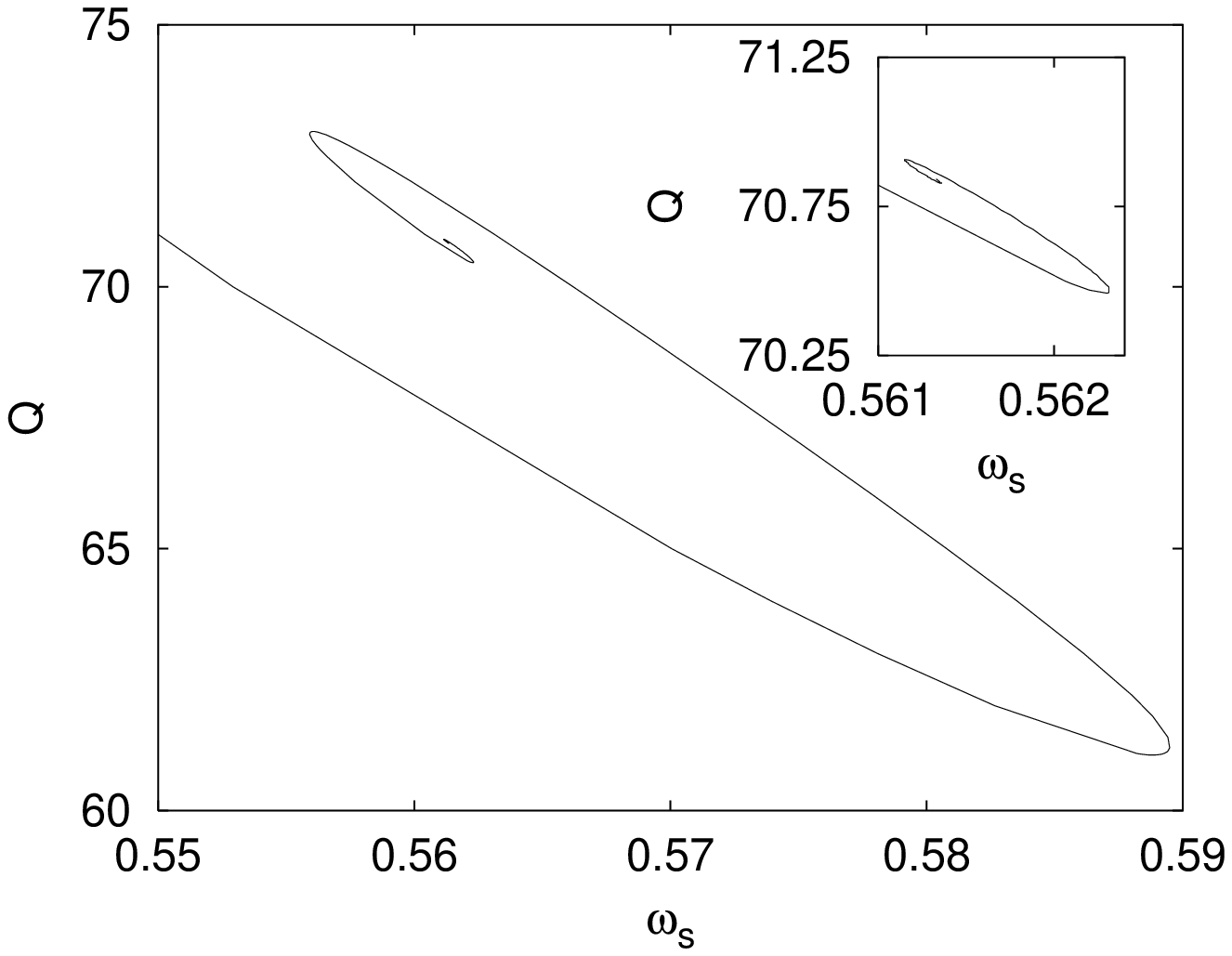}
}}}
\caption{
%5
The charge $Q$ is shown as a function of the frequency $\omega_s$
for fundamental boson stars ($k=0$) 
in the full range of existence (left),
and in the frequency range of the spiral (right),
for the gravitational coupling constant $\kappa=0.2$.
}
\label{qvsos_col}
\end{figure}

To demonstrate the effects of gravity on the
spherically symmetric solutions, we exhibit in Fig.~\ref{qvsos_col}
the charge $Q$ as a function of the
frequency $\omega_s$ for the fundamental boson star solutions
at a given value of the gravitational coupling constant $\kappa$.

For solutions in curved space the frequency $\omega_s$ 
is also bounded from above by
$\omega_{\rm max}$, Eq.~(\ref{cond1}),
since the scalar field exhibits asymptotically an exponential fall-off
only for $\omega_s < \omega_{\rm max}$.
However, for the smaller values of $\omega_s$ a new phenomenon occurs,
as compared to flat space \cite{lee-bs},
where the solutions approach monotonically the limiting lower
value $\omega_{\rm min}$, Eq.~(\ref{cond2}).
Denoting the minimal value of the frequency for boson stars
by $\omega_0(\kappa)$, we observe that it
differs from $\omega_{\rm min}$, except for a single value
of the gravitational coupling $\kappa$.
For larger values of $\kappa$, $\omega_0(\kappa) > \omega_{\rm min}$,
while for smaller values of $\kappa$, $\omega_0(\kappa) < \omega_{\rm min}$.
(Note, that the latter case does not occur for a potential
with degenerate minima, since there $\omega_{\rm min}=0$.)
Moreover, in the presence of gravity the solutions do not
approach monotonically the minimal value $\omega_0(\kappa)$.
Instead one observes an inspiralling of the boson star solutions
towards a limiting solution at the center of the spiral
at a frequency $\omega_{\rm lim}(\kappa) > \omega_0(\kappa)$ \cite{lee-bs}.

\begin{figure}[h!]
\parbox{\textwidth}
{\centerline{
\mbox{
\epsfysize=10.0cm
\includegraphics[width=70mm,angle=0,keepaspectratio]{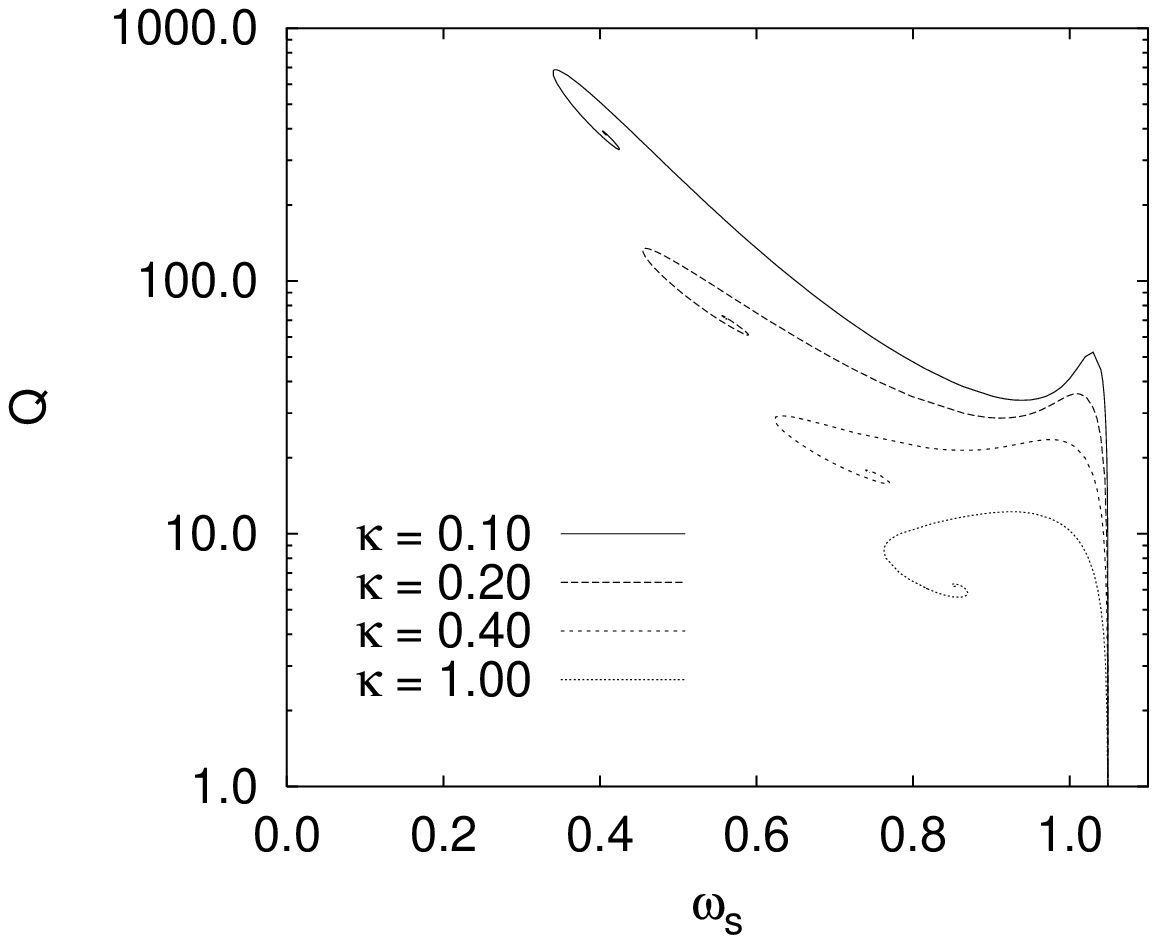}
\includegraphics[width=70mm,angle=0,keepaspectratio]{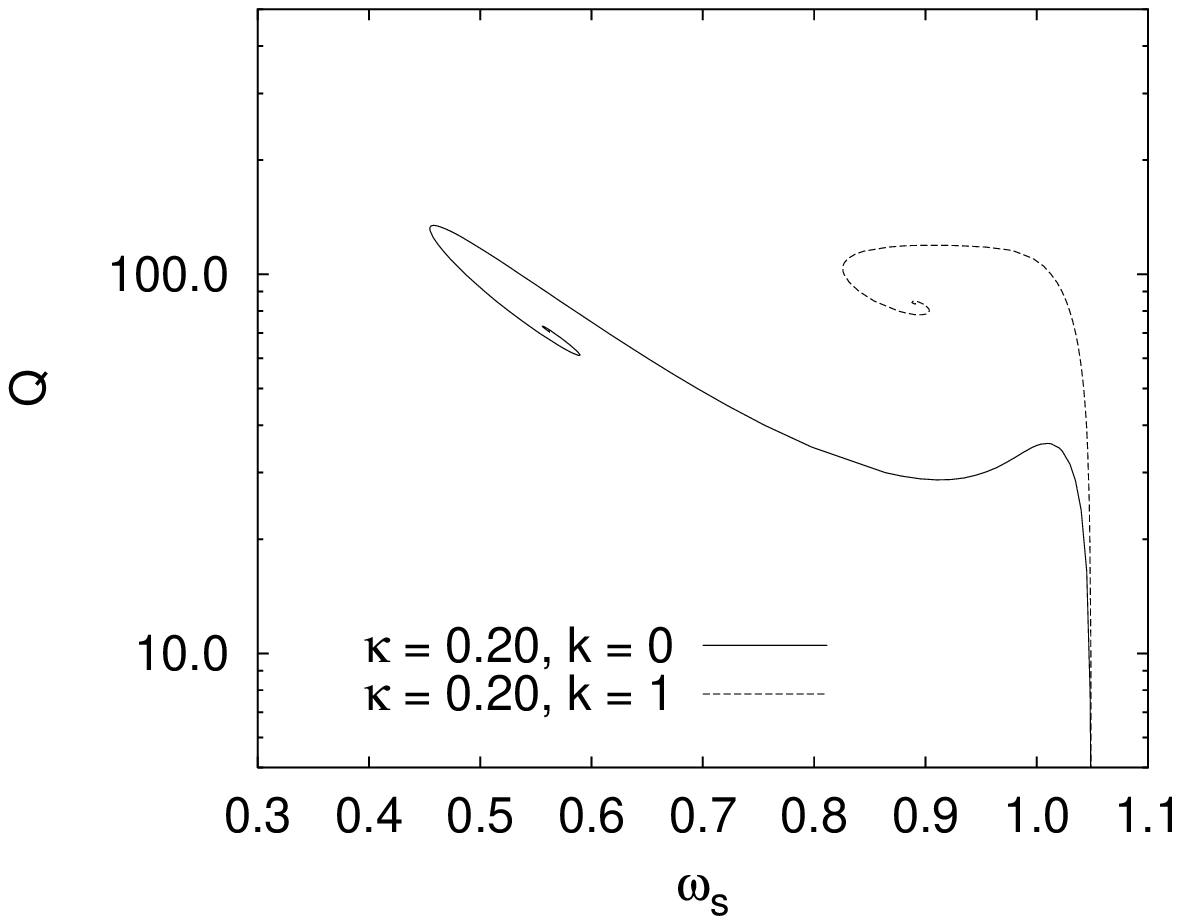}
}}}
\caption{
%6
The charge $Q$ is shown as a function of the frequency $\omega_s$
for fundamental boson stars ($k=0$)
for the values of the gravitational coupling constant 
$\kappa=0.1$, 0.2, 0.4, 1.0 (left),
and for their first radial excitations ($k=1$) 
for $\kappa=0.2$ (right).
}
\label{Qvsomega01}
\end{figure}

In Fig.~\ref{Qvsomega01} we exhibit the charge $Q$ as a function of the
frequency $\omega_s$, for the fundamental boson star solution ($k=0$)
for several values of the gravitational coupling constant $\kappa$,
and also for the first radial excitation ($k=1$)
for $\kappa=0.2$.
While the presence of a spiral is a genuine property of boson stars, 
the location and the size of the spiral depend on
the gravitational coupling strength $\kappa$ and on the node number $k$.

\begin{figure}[h!]
\parbox{\textwidth}
{\centerline{
\mbox{
\epsfysize=10.0cm
\includegraphics[width=70mm,angle=0,keepaspectratio]{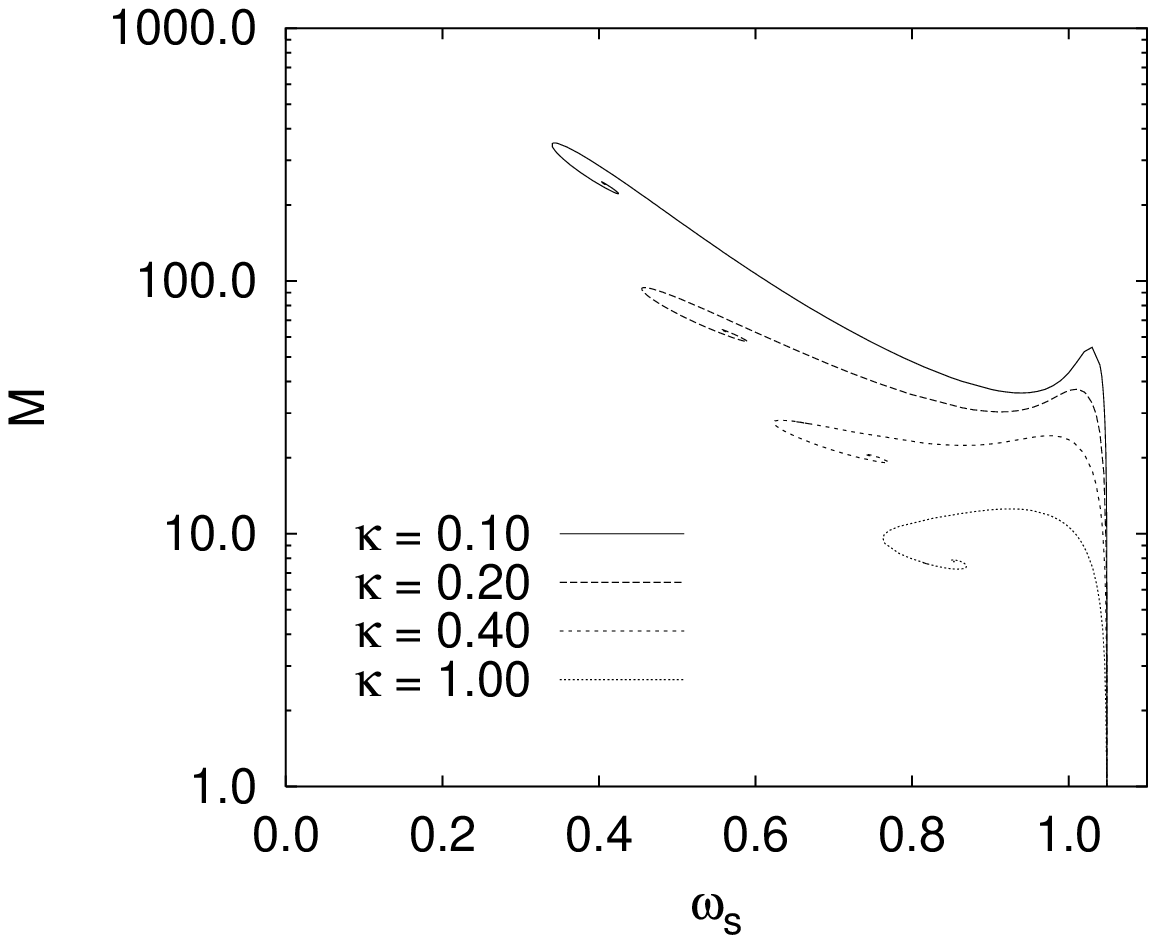}
\includegraphics[width=70mm,angle=0,keepaspectratio]{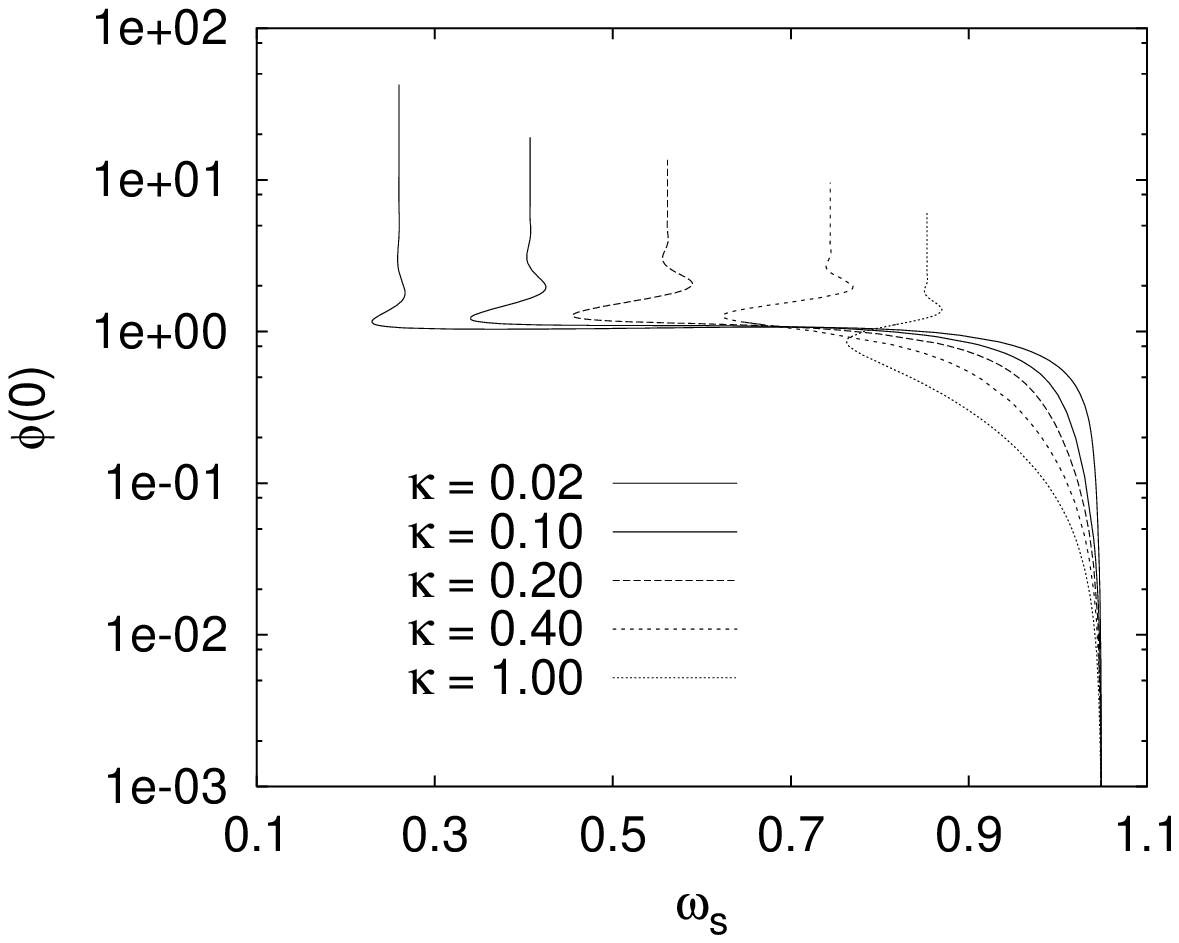}
}}}
\caption{
%7
The mass $M$ (left) 
and the value of the scalar field at the origin $\f(0)$ (right)
are shown as functions of the frequency $\omega_s$
for fundamental boson stars ($k=0$)
for the values of the gravitational coupling constant
$\kappa=0.1$, 0.2, 0.4, 1 (left) 
resp.~$\kappa=0.02$, 0.1, 0.2, 0.4, 1 (right).
}
\label{mvsos_col}
\end{figure}

The mass $M$ has an analogous dependence on the frequency $\omega_s$
as the charge $Q$, as seen in Fig.~\ref{mvsos_col},
where we also exhibit the frequency dependence of
the value of the scalar field at the origin $\f(0)$. 
%The endpoints of $\f(0)$ correspond to the finite values $\f_{\rm lim}(0)$,
%assumed at the limiting solutions at the centers of the spirals.
The endpoints of $\f(0)$ in the figure correspond to the 
numerical values obtained closest to the centers of the spirals.
We cannot decide numerically, however, whether the limiting solutions
have indeed finite values $\f_{\rm lim}(0)$ \cite{footlim}.

\begin{figure}[h!]
\parbox{\textwidth}
{\centerline{
\mbox{
\epsfysize=10.0cm
\includegraphics[width=70mm,angle=0,keepaspectratio]{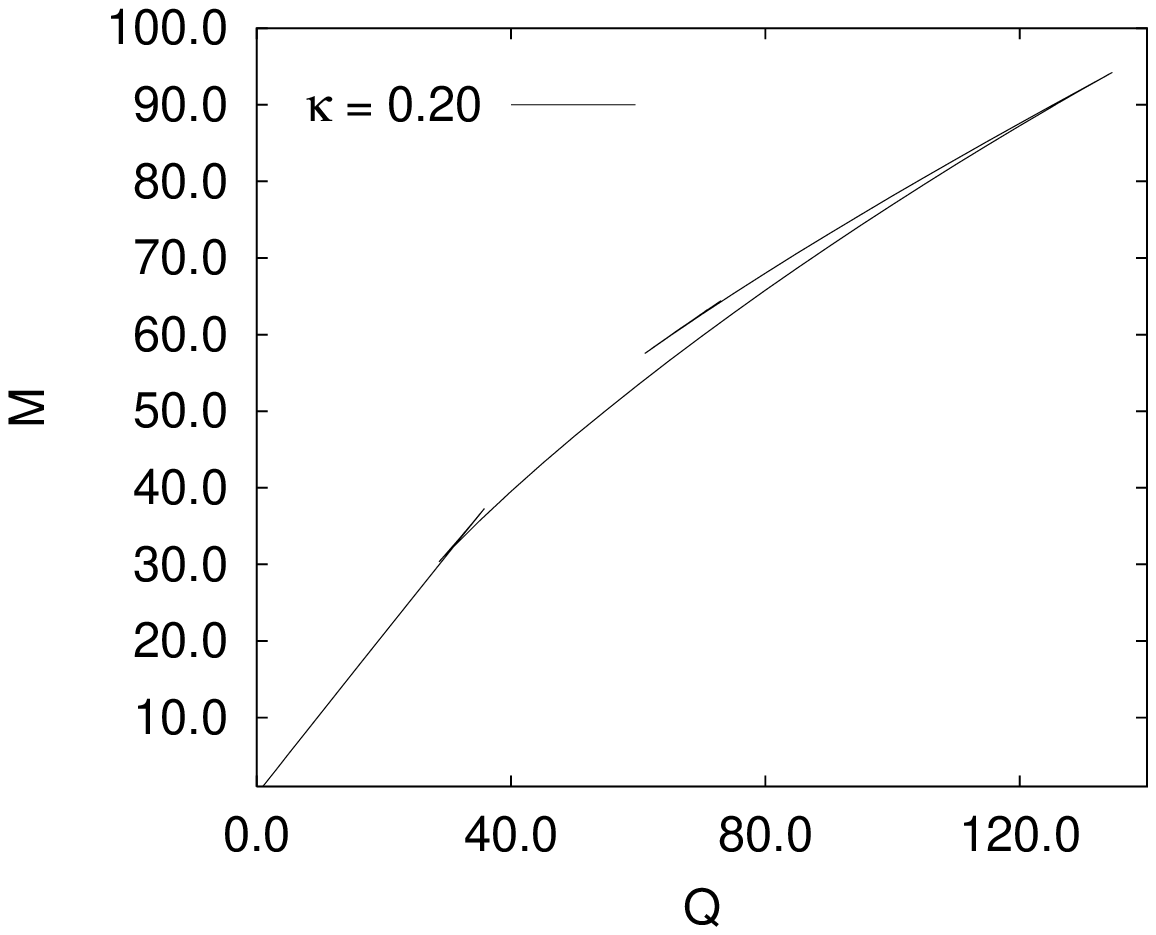}
\includegraphics[width=70mm,angle=0,keepaspectratio]{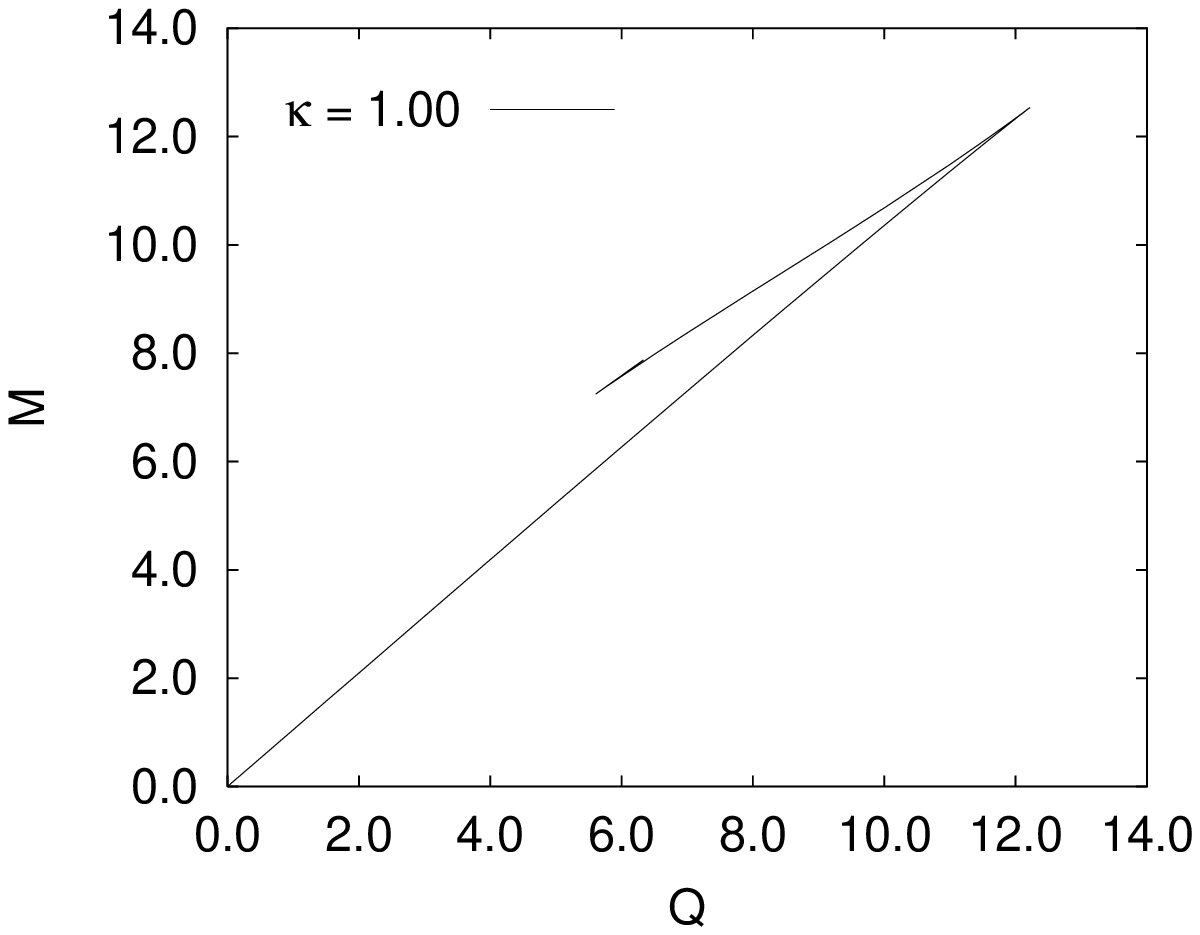}
}}}
\caption{
%8
The mass $M$ 
is shown as a function of the charge $Q$
for fundamental boson stars ($k=0$)
for the values of the gravitational coupling constant
$\kappa=0.2$ (left) and $\kappa=1$ (right).
}
\label{mvsq_col}
\end{figure}

When the mass $M$ is considered as a function of the charge $Q$,
we observe a cusp structure \cite{lee-bs}, 
as illustrated in Fig.~\ref{mvsq_col}.
For smaller values of $\kappa$ (e.g.,~$\kappa=0.2$)
we observe two sets of cusps. The first set of cusps
is related to the single cusp present in flat space, which occurs at the
minimal value of the charge $Q_{\rm min}$, 
where the upper and lower branch of the $Q$-ball merge.
But whereas the upper branch extends infinitely far in flat space,
it extends in curved space only up to a second cusp, reached at 
a second critical value of the charge, where this upper branch
merges with a third branch extending down to zero.
For larger values of $\kappa$ (e.g.,~$\kappa=1$)
this set of two cusps, being a
remainder of the flat space solutions, is no longer present.

The second set of cusps is related to the spirals, present only
in curved space.
Labelling these cusps of the spirals consecutively $N=1,2,...$,
we exhibit the charge $Q$, the mass $M$,
and the frequency $\omega_s$ of the boson star solutions at these cusps
in Table~\ref{table_col_alp},
for several values of the gravitational coupling constant
$\kappa$.

\begin{table}[h!]
%\begin{center}
\parbox{\textwidth}
{\centerline{
\mbox{
\begin{tabular}{|c|c|c|c|c|}
\hline
$N$ & $Q_N$ & $M$ & $\omega_s$ \\
\hline
\hline
$1$ & $685.5000$ & $351.6320$ & $0.34338$ \\
$2$ & $331.3000$ & $221.3990$ & $0.42424$ \\
$3$ & $391.9200$ & $246.5870$ & $0.40352$ \\
$4$ & $378.2500$ & $241.0480$ & $0.40776$ \\
$5$ & $380.8700$ & $242.1160$ & $0.40693$ \\
$6$ & $380.4000$ & $241.9250$ & $0.40708$ \\
\hline
\hline
$N$ & $Q_N$ & $M$ & $\omega_s$ \\
\hline
\hline
$1$ & $29.3040$ & $28.1273$ & $0.63000$ \\
$2$ & $15.7880$ & $19.1084$ & $0.76800$ \\
$3$ & $17.8630$ & $20.6836$ & $0.74018$ \\
$4$ & $17.4580$ & $20.3833$ & $0.74497$ \\
$5$ & $17.5280$ & $20.4361$ & $0.74416$ \\
$6$ & $17.5170$ & $20.4273$ & $0.74429$ \\
\hline
\end{tabular}
\begin{tabular}{|c|c|c|c|c|}
\hline
$N$ & $Q_N$ & $M$ & $\omega_s$ \\
\hline
\hline
$1$ & $119.2400$ & $122.9620$ & $0.89402$ \\
$2$ & \ $78.1580$ & \ $88.3579$ & $0.89343$ \\
$3$ & \ $84.8980$ & \ $94.4262$ & $0.89000$ \\
$4$ & \ $83.4960$ & \ $93.1796$ & $0.89067$ \\
$5$ & \ $83.7470$ & \ $93.4033$ & $0.89047$ \\
$6$ & \ $83.7090$ & \ $93.3694$ & $0.89048$ \\
\hline
\hline
$N$ & $Q_N$ & $M$ & $\omega_s$ \\
\hline
\hline
$1$ & $12.2180$ & $12.5360$ & $0.93000$ \\
$2$ & \ $5.6059$ & $7.24552$ & $0.86000$ \\
$3$ & \ $6.3314$ & $7.87359$ & $0.85305$ \\
$4$ & \ $6.1911$ & $7.75446$ & $0.85290$ \\
$5$ & \ $6.2148$ & $7.77456$ & $0.85304$ \\
$6$ & \ $6.2108$ & $7.77104$ & $0.85302$ \\
\hline
\end{tabular}
%\end{center}
}}}
\caption{Physical characteristics of the first six cusps of the fundamental
boson star solutions ($k=0$)
for gravitational coupling constants $\kappa=0.1$ (upper left),
$\kappa=0.2$ (upper right),
$\kappa=0.4$ (lower left),
$\kappa=1$ (lower right).}
\label{table_col_alp}
\end{table}

\begin{figure}[h!]
\parbox{\textwidth}
{\centerline{
\mbox{
\epsfysize=10.0cm
\includegraphics[width=70mm,angle=0,keepaspectratio]{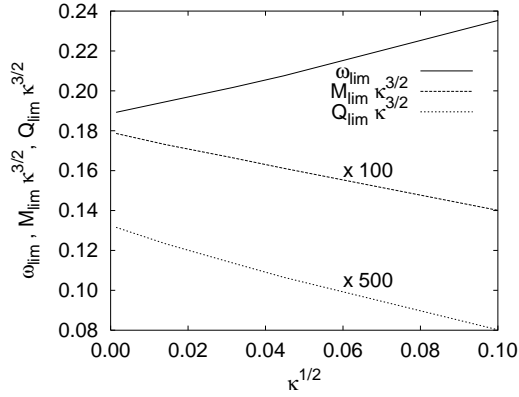}
}}}
\caption{
%8-1
The dependence of the frequency $\omega_{\rm lim}$,
the mass $M_{\rm lim}$ and the charge $Q_{\rm lim}$ 
of the limiting solutions at the centers of the spirals
on the gravitational coupling constant $\kappa$
for fundamental boson stars ($k=0$)
for small values of $\kappa$.
}
\label{kappalim}
\end{figure}

Focussing on the limiting solutions 
at the centers of the spirals, we exhibit in
Table~\ref{tabl} the limiting values of the frequency $\omega_{\rm lim}$,
the charge $Q_{\rm lim}$ and the mass $M_{\rm lim}$ for several
values of the gravitational coupling constant $\kappa$.
For small gravitational coupling we extract the following $\kappa$-dependence,
\begin{eqnarray}
\omega_{\rm lim} 
 &=& c_0^\omega + c_1^\omega \kappa^{1/2}  + c_2^\omega \kappa + O(\kappa^{3/2})
 \ ,  \nonumber\\
Q_{\rm lim} \kappa^{3/2}
 &=& c_0^Q + c_1^Q \kappa^{1/2}  + c_2^Q \kappa + O(\kappa^{3/2})
 \ ,   \nonumber\\
M_{\rm lim} \kappa^{3/2}
 &=& c_0^M + c_1^M \kappa^{1/2}  + c_2^M \kappa + O(\kappa^{3/2})
 \ ,
\end{eqnarray}
illustrated in Fig.~\ref{kappalim}. 
For small gravitational coupling we thus obtain a different
$\kappa$-dependence, as compared to \cite{lee-bs}.
In particular, in the limit $\kappa \rightarrow 0$
the limiting frequency $\omega_{\rm lim}$ assumes a finite value,
$0 < \omega_{\rm lim}(0) < \omega_{\rm min}$.
We attribute this difference to the fact, that for a potential
with degenerate minima, as employed in \cite{lee-bs},
$\omega_{\rm min}=0$. 
\begin{table}[h!]
\begin{center}
\begin{tabular}{|c|c|c|c|}
\hline
$\kappa$ & $\omega_{\lim}$ & $Q_{\lim}$ & $M_{\lim}$  \\
\hline
\hline
$1.0000$ & $0.85302$ & \ \ $6.2114$ & \ \ $7.7718$ \\
$0.4000$ & $0.74427$ & \ $17.5180$ & \ $20.4285$  \\
$0.2000$ & $0.56135$ & \ $70.8330$ & \ $63.2038$  \\
$0.1000$ & $0.40707$ & $380.4600$ & $241.9480$  \\
$0.0200$ & $0.25981$ & $1.136 \cdot 10^{4}$ & $4433.790$  \\
$0.0100$ & $0.23529$ & $4.011 \cdot 10^{4}$ & $1.401 \cdot 10^{4}$  \\
$0.0020$ & $0.20754$ & $5.954 \cdot 10^{5}$ & $1.802 \cdot 10^{5}$  \\
$0.0010$ & $0.20172$ & $1.797 \cdot 10^{6}$ & $5.262 \cdot 10^{5}$  \\
$0.0002$ & $0.19451$ & $2.181 \cdot 10^{7}$ & $6.119 \cdot 10^{6}$  \\
$0.2 \cdot 10^{-5} $ & $0.18924$ & $2.326 \cdot 10^{10}$ & $6.316 \cdot 10^{9}$
 \\
\hline
\end{tabular}
\end{center}
\caption{Physical characteristics of the limiting solutions
at the centers of the spirals
for the fundamental boson stars ($k=0$)
for a set of decreasing values of the
gravitational coupling constant $\kappa$.
}
\label{tabl}
\end{table}

\begin{figure}[h!]
\parbox{\textwidth}
{\centerline{
\mbox{
\epsfysize=10.0cm
\includegraphics[width=70mm,angle=0,keepaspectratio]{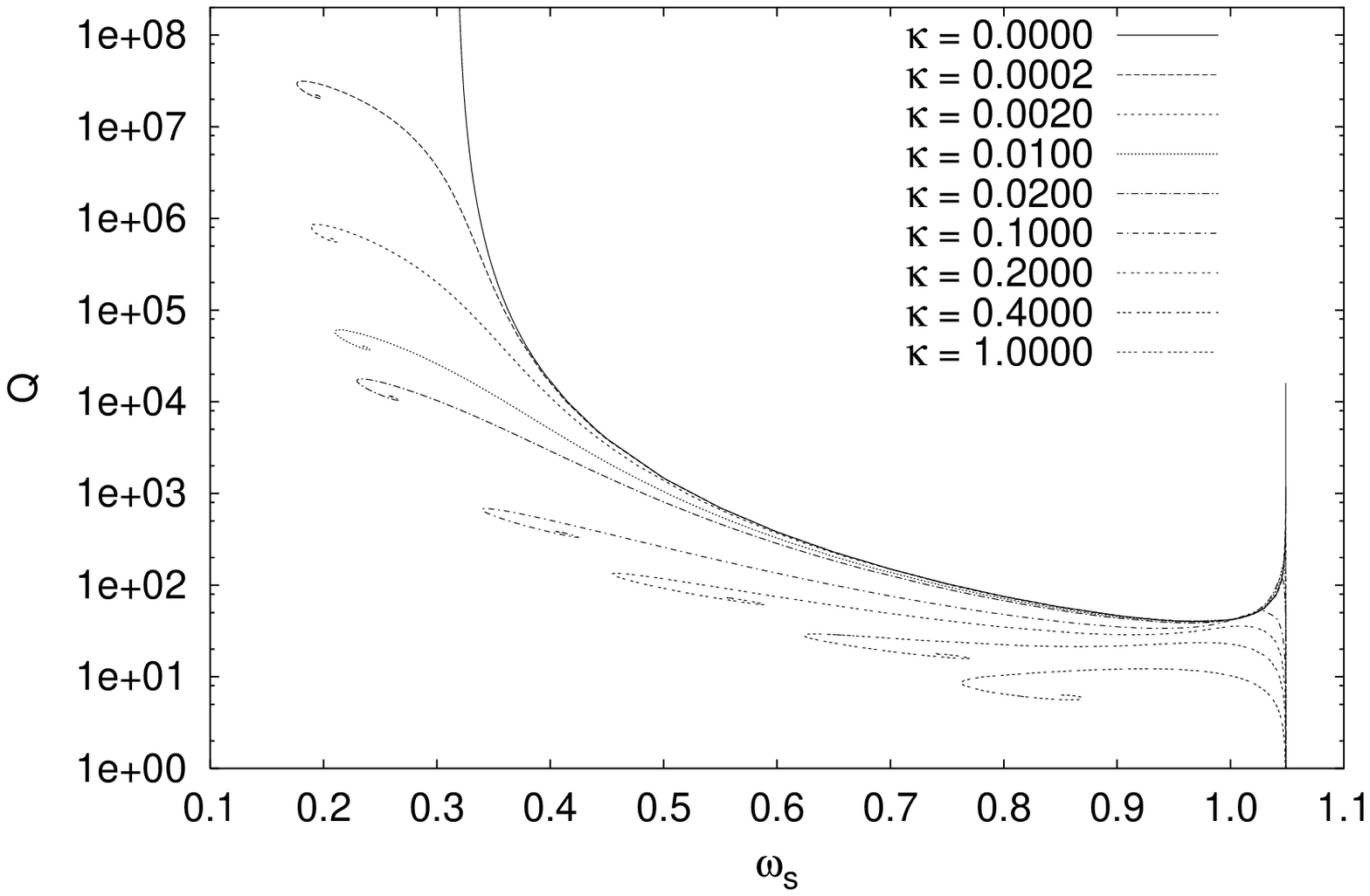}
\includegraphics[width=70mm,angle=0,keepaspectratio]{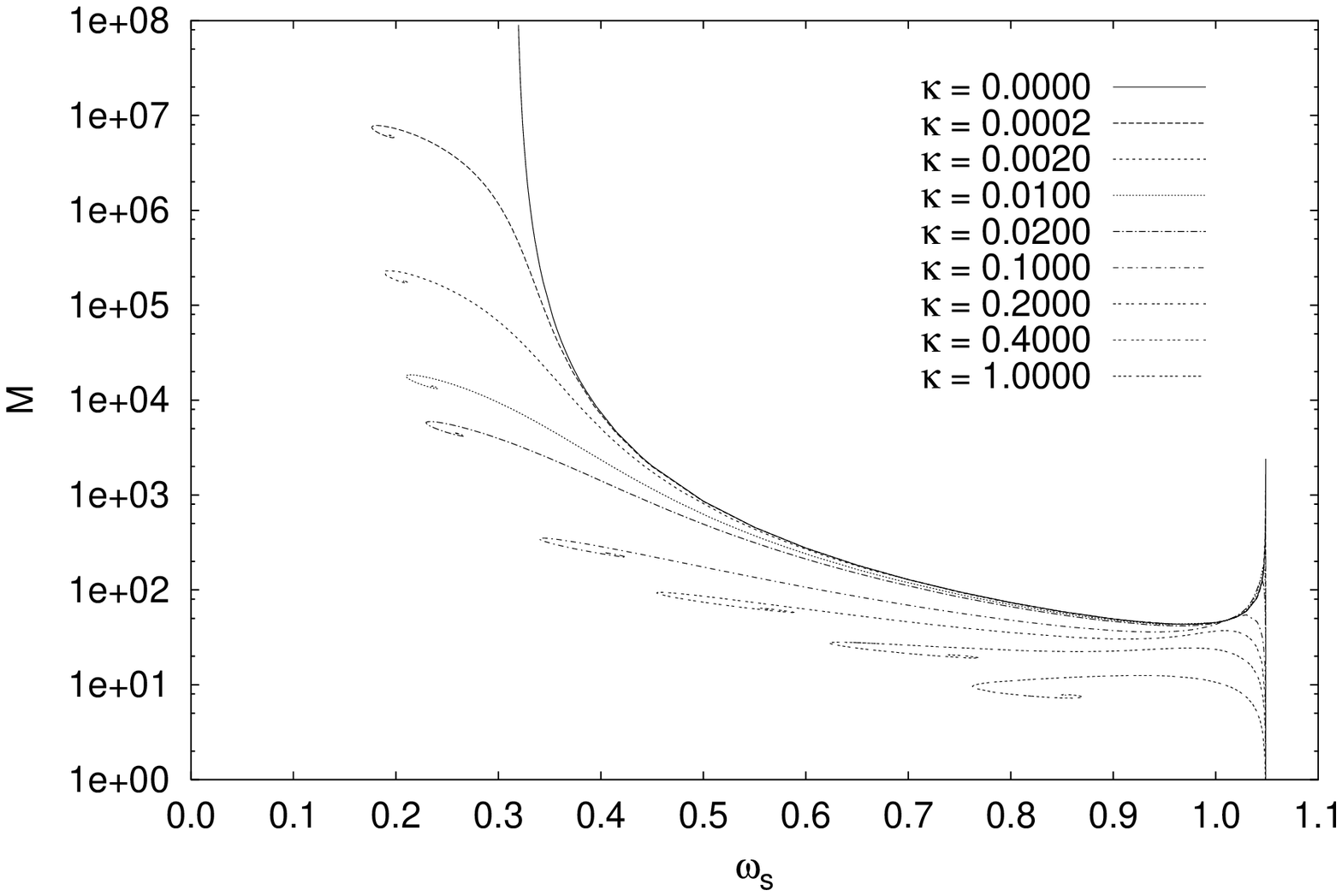}
}}}
\caption{
%9
The charge $Q$ (left) and the mass $M$ (right)
are shown as functions of the frequency $\omega_s$
for fundamental boson stars ($k=0$)
for a set of decreasing values of the gravitational coupling constant
$\kappa$. Also shown are the limiting flat space values.
}
\label{Qlim}
\end{figure}

To address the $\kappa$-dependence of the domain
of existence of the (fundamental) boson star solutions, 
corresponding to the interval $[\omega_0(\kappa), \omega_{\rm max}]$,
we exhibit in Fig.~\ref{Qlim} the charge $Q$ 
as a function of the frequency $\omega_s$ for 
a large set of values of
the gravitational coupling constant $\kappa$,
including the flat space limit.
Interestingly, for values of the frequency $\omega_s$ close to
$\omega_{\rm max}$, the flat space values are not approached
monotonically from below with decreasing $\kappa$.
Here a weak coupling to gravity can lead to an increase of the mass of the
boson star solutions.

\begin{figure}[h!]
\parbox{\textwidth}
{\centerline{
\mbox{
\epsfysize=10.0cm
\includegraphics[width=70mm,angle=0,keepaspectratio]{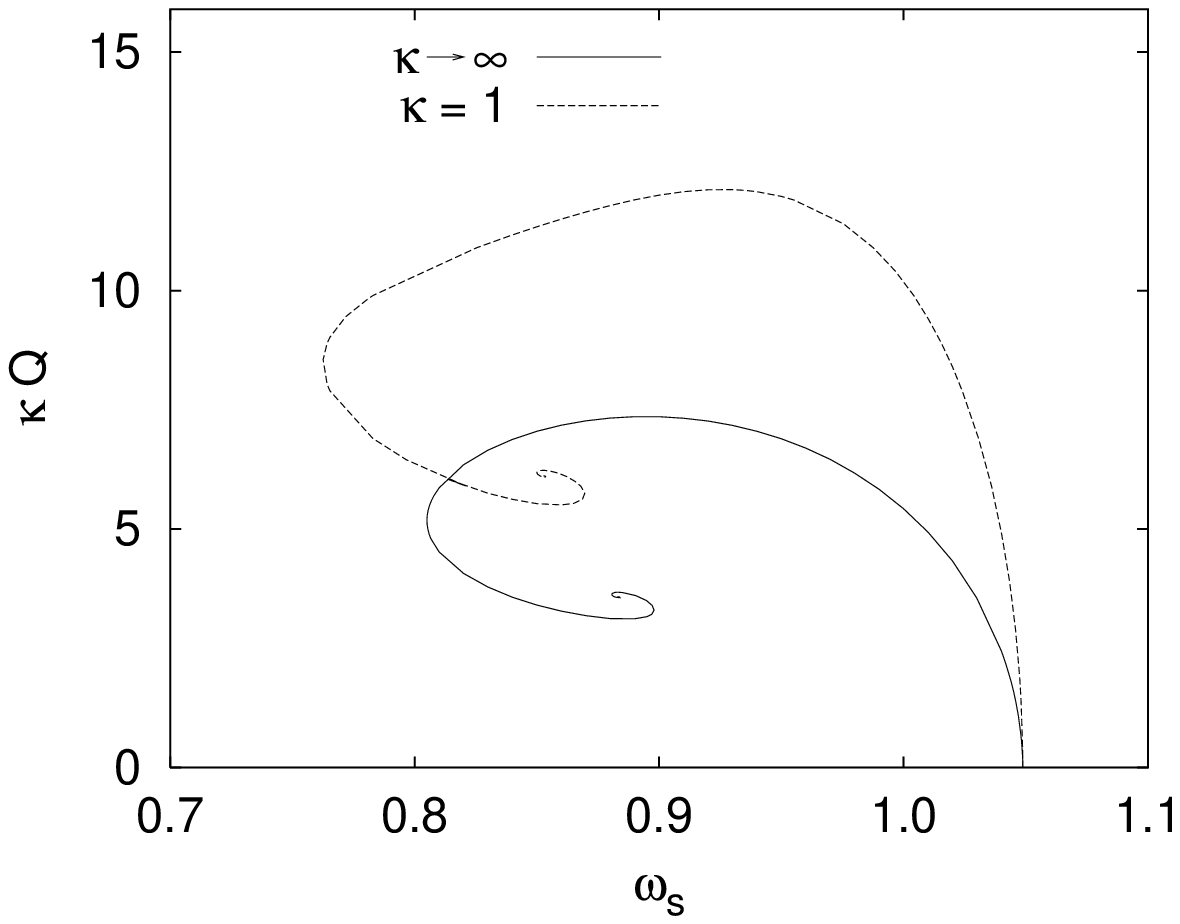}
\includegraphics[width=70mm,angle=0,keepaspectratio]{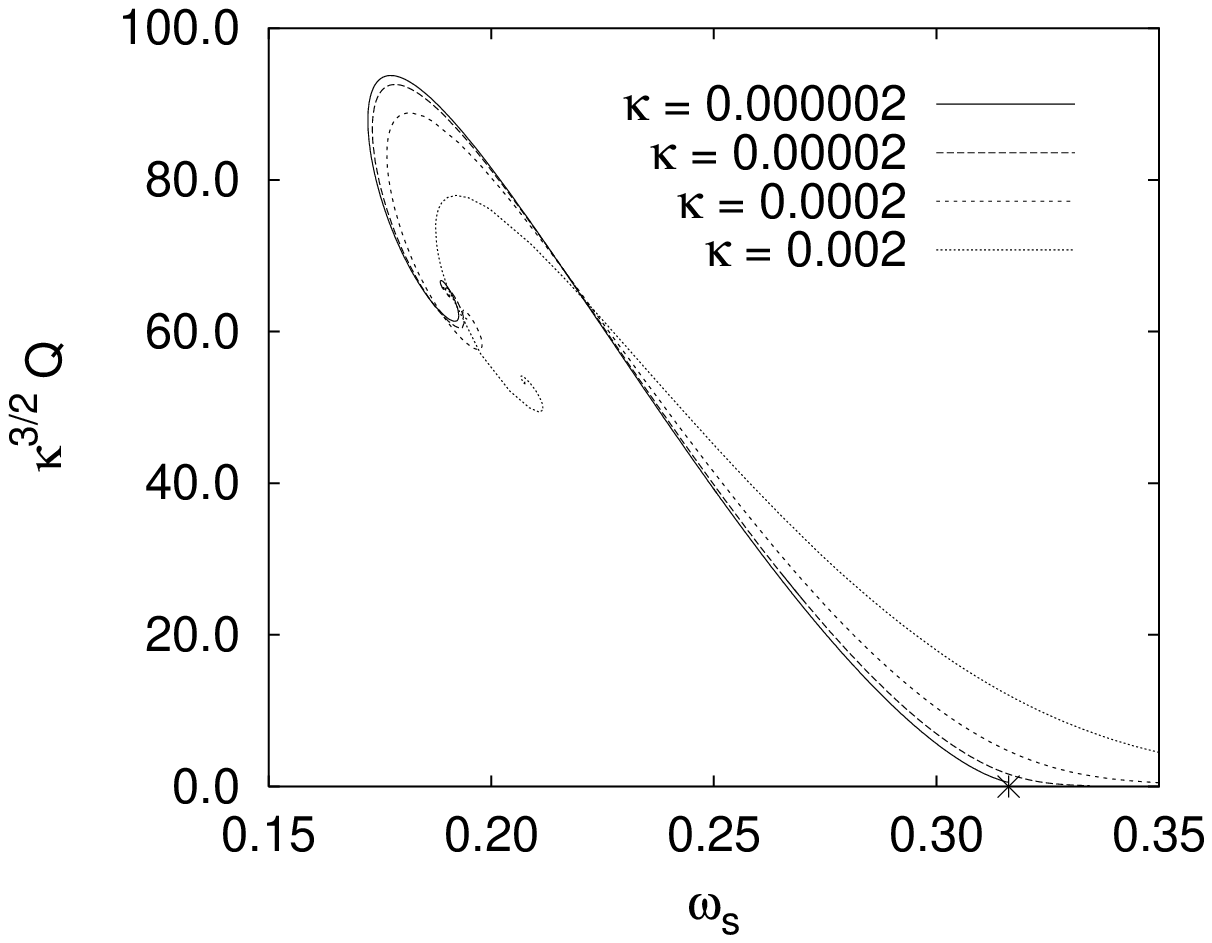}
}}}
\caption{
%9-1
Left:
The scaled charge $\kappa Q$ 
is shown as a function of the frequency $\omega_s$
for fundamental boson stars ($k=0$) 
in the limit $\kappa \rightarrow \infty$.
For comparison the scaled charge $\kappa Q$ 
is also shown for $\kappa = 1$.
resp.~$\kappa \rightarrow 0$ (right).
Right:
The scaled charge $\kappa^{3/2} Q$ 
is shown as a function of the frequency $\omega_s$
for fundamental boson stars ($k=0$) 
for $\kappa = 0.002$, 0.0002, 0.00002, and 0.000002.
The asterisk marks $\omega_{\rm min}$.
}
\label{Qinfty}
\end{figure}

As $\kappa$ increases, $\omega_0(\kappa)$ tends to a finite 
value $\omega_0(\infty)$, smaller than $\omega_{\rm max}$.
Fixing $\omega_s \in [\omega_0(\infty),\omega_{\rm max}]$ 
we observe that the scalar field scales like $1/\sqrt{\kappa}$
for large values of $\kappa$.
In order to obtain the solutions 
in the limit $\kappa \rightarrow \infty$, we therefore introduce the scaled
scalar field $\hat{\phi}(r)=\sqrt{\kappa} \phi(r)$. 
Substituting $\phi(r)$ in the field equation and taking the
limit $\kappa \rightarrow \infty$, we find that all terms non-linear 
in $\hat{\phi}(r)$ vanish. Similarly, in the Einstein equations 
all terms of higher than second order in $\hat{\phi}(r)$ vanish and
the dependence on $\kappa$ cancels.
Thus we end up with a set of differential equations identical to
the original one, except that $\kappa=1$ and 
$U(\hat{\phi})= \lambda b \hat{\phi}^2$.
Solving this set of differential equations for fixed $\lambda b$ and 
varying $\omega_s$, we again find a spiral pattern for the 
(scaled) charge $\hat{Q} = \kappa Q$, shown in Fig.~\ref{Qinfty}.
This analysis yields
$$ 
\omega_0(\infty) = 0.805\ , 
\ \ \ \ 
\omega_{\lim}(\infty) = 0.883\ . 
$$

Considering now the domain of existence in the limit 
$\kappa \rightarrow 0$, we note
that we must distinguish two intervals here,
$[\omega_0(\kappa), \omega_{\rm min}]$ and
$[\omega_{\rm min}, \omega_{\rm max}]$.
We observe from Fig.~\ref{Qlim},
that the limiting flat space values of the charge $Q$ are approached in a
continuously increasing interval,
extending to the full interval $[\omega_{\rm min}, \omega_{\rm max}]$
in the limit $\kappa \rightarrow 0$.
In the interval $[\omega_0(0), \omega_{\rm min}]$
there are no flat space solutions,
possessing finite charge or mass.
As noted above, the limiting values 
of the charge $Q_{\rm lim}$ and the mass $M_{\rm lim}$ 
at the centers of the spirals
diverge with $\kappa^{-3/2}$,
when $\kappa \rightarrow 0$.
We therefore determine the properties of the solutions 
in the limit $\kappa \rightarrow 0$
in the interval $[\omega_0(0), \omega_{\rm min}]$
by scaling the charge and the mass with $\kappa^{3/2}$
for a sequence of solutions, corresponding to decreasing values of $\kappa$.
Convergence towards the limiting values of the scaled charge is demonstrated
in Fig.~\ref{Qinfty} \cite{footzero}.
Thus also in the limit $\kappa \rightarrow 0$ we find a spiral pattern for the
(scaled) charge $\hat{Q} = \kappa^{3/2} Q$.
This analysis yields
$$
\omega_0(0) = 0.172\ ,
\ \ \ \
\omega_{\lim}(0) = 0.189 \ .
$$
We remark that for a potential with degenerate minima, 
as employed in \cite{lee-bs}, only the single interval
$[\omega_{\rm min}, \omega_{\rm max}]$ needs to be considered,
since $\omega_{\rm min}=0$, implying apparently
$\omega_0(0) = \omega_{\lim}(0) = \omega_{\rm min} =0$.

\section{Rotating Axially Symmetric Solutions}\label{c4}

Rotating axially symmetric solutions are obtained, when
$n \ne 0$.
We solve the set of coupled non-linear
elliptic partial differential equations, given in Appendix \ref{dgl_sys2},
numerically \cite{schoen},
subject to the above boundary conditions, Eqs.~(\ref{bc3})-(\ref{bc6}),
employing the compactified radial coordinate, Eq.~(\ref{rcomp}).
The numerical calculations are based on the Newton-Raphson method.
The equations are discretized on a non-equidistant grid in
$\bar r$ and  $\theta$.
Typical grids used have sizes $100 \times 20$,
covering the integration region
$0\leq \bar r\leq 1$ and $0\leq\theta\leq\pi/2$.

\boldmath
\subsection{Rotating $Q$-balls}
\unboldmath

The existence of rotating $Q$-balls has been shown by
Volkov and W\"ohnert \cite{volkov}.
Based on the ansatz (\ref{ansatzp}) for the scalar field $\Phi$ \cite{schunck},
one obtains for rotating $Q$-ball solutions the field equation \cite{foot1}
\begin{eqnarray}
\label{eom3d}
\left( \frac{\partial^2} {\partial r^2} + \frac{2} {r} \frac{\partial}
{\partial r} + \frac{1} {r^2} \frac{\partial^2} {\partial \theta^2}
+ \frac{\cos\theta} {r^2 \sin\theta} \frac{\partial} {\partial
\theta} - \frac{n^2} {r^2 \sin^2\theta} + \omega_s^2 \right) \phi &
= & \frac{1}{2} \frac{d U(\phi)}{d \phi} \ , \phantom{abcd}
\end{eqnarray}
the mass \cite{volkov}
\begin{equation}
\label{erot}
M = 2 \pi \int_0^\infty \int_0^\pi 
\left(\omega_s^2 \phi^2 + (\partial_r \phi)^2 +
\frac{1} {r^2} (\partial_\theta \phi)^2 + \frac{n^2\phi^2} {r^2
\sin^2 \theta} + U(\phi) \right) 
r^2 \, d r \, \sin\theta d \theta
\ , 
\end{equation}
and the charge
\begin{equation}
Q= 4 \pi \omega_s \int_0^{\infty}\int _0^{\pi} 
 \f^2 \, r^2 dr \, \sin \theta \, d\t \ 
\ . \label{3} \end{equation}
Their angular momentum satisfies the quantization relation
$J=nQ$.

In their pioneering study \cite{volkov}
Volkov and W\"ohnert 
show, that for a given value of $n$ there are two types of solutions,
possessing different parity.
They exhibit examples of fundamental rotating $Q$-balls
with quantum numbers $n=1-3$, and both even and odd parity.
They do not study the frequency dependence of 
rotating $Q$-ball solutions, however.

\begin{figure}[h!]
\parbox{\textwidth}
{\centerline{
\mbox{
\epsfysize=10.0cm
\includegraphics[width=70mm,angle=0,keepaspectratio]{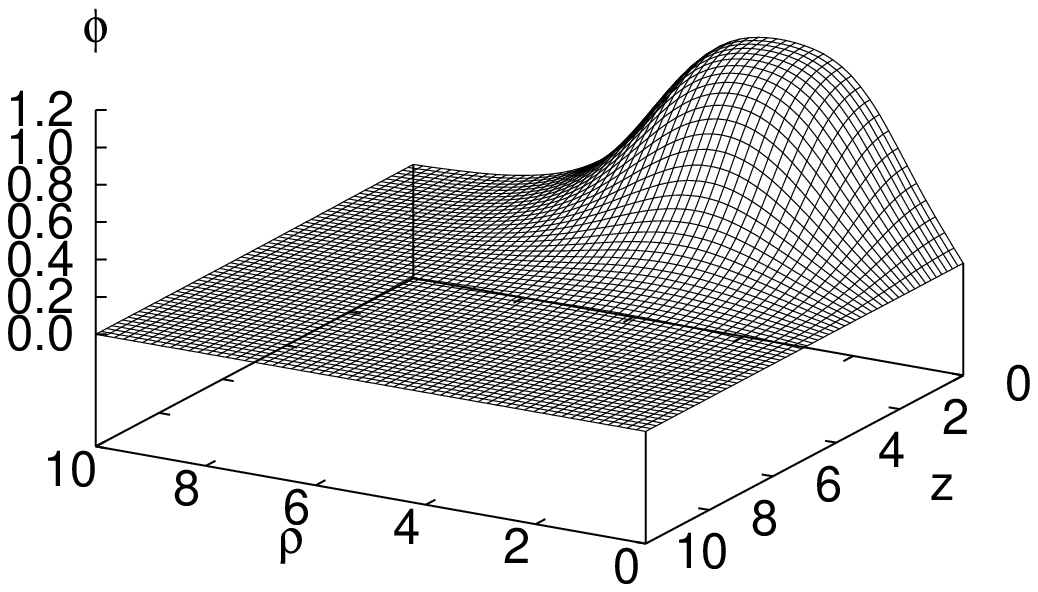}
\includegraphics[width=70mm,angle=0,keepaspectratio]{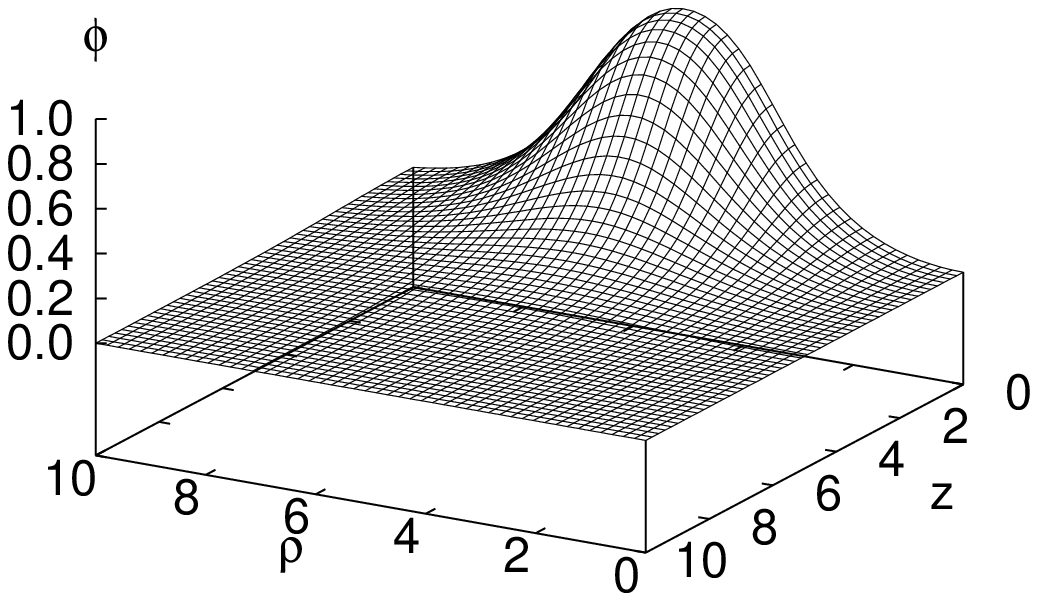}
}}}
\caption{
%10
The scalar field $\f(r,\t)$ for fundamental rotating $Q$-balls
with charge $Q=410$ and quantum number $n=1$ (left) and $n=2$ (right)
is shown as a function of the coordinates $\rho=r \sin \theta$
and $z= r \cos\theta$.
}
\label{Q410_n2}
\end{figure}

\begin{figure}[h!]
\parbox{\textwidth}
{\centerline{
\mbox{
\epsfysize=10.0cm
\includegraphics[width=70mm,angle=0,keepaspectratio]{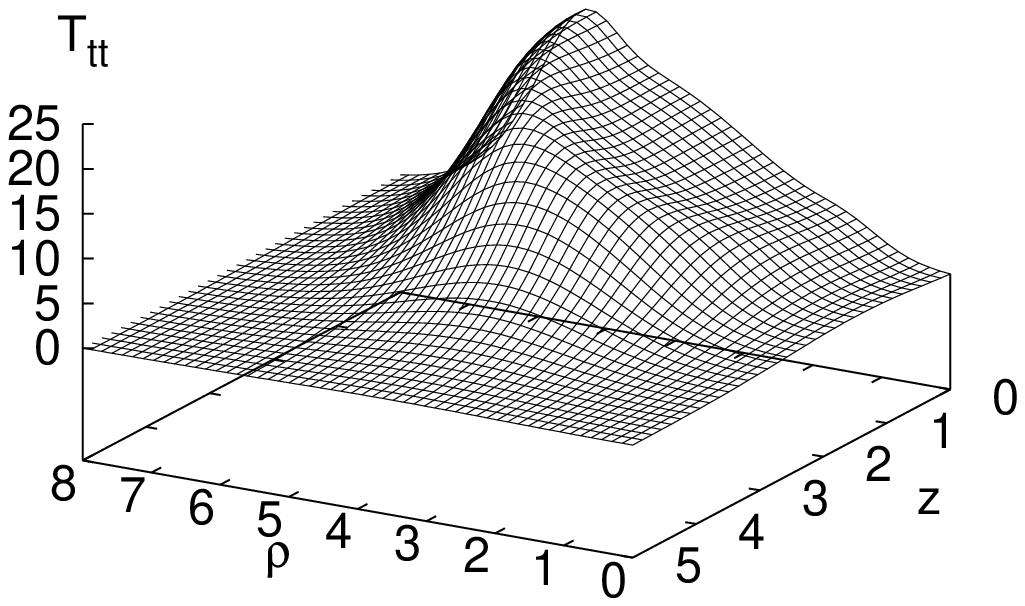}
\includegraphics[width=70mm,angle=0,keepaspectratio]{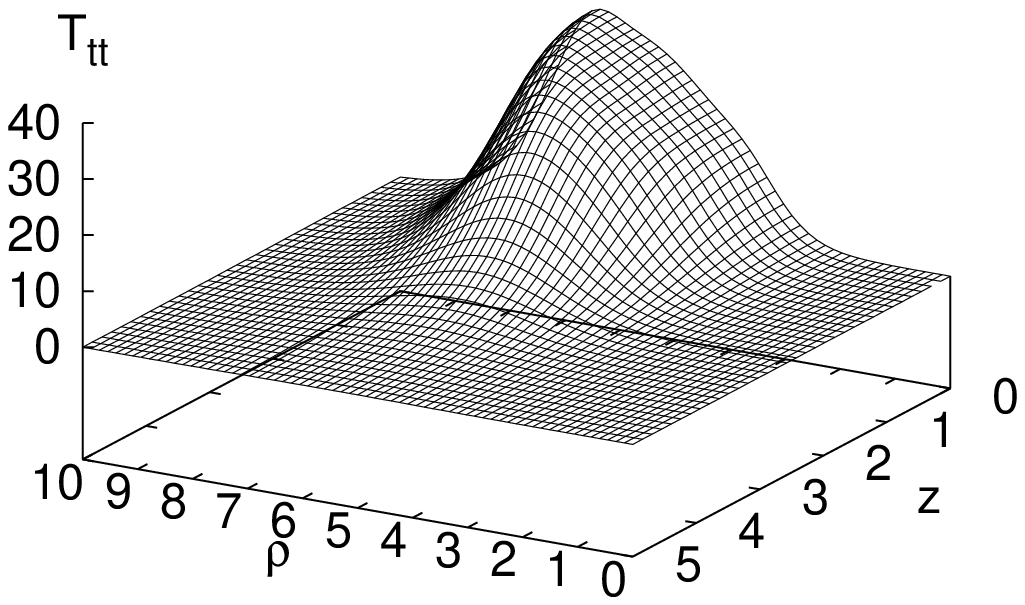}
}}}
\caption{
%11
The energy density $T_{tt}$ for fundamental rotating $Q$-balls
with charge $Q=410$ and quantum number $n=1$ (left) and $n=2$ (right)
is shown as a function of the coordinates $\rho=r \sin \theta$
and $z= r \cos\theta$.
}
\label{T410_n2}
\end{figure}

We illustrate the scalar field $\f(r,\theta)$ and the energy density
$T_{tt}(r,\theta)$ of fundamental rotating $Q$-balls with charge $Q=410$,
quantum numbers $n=1$ and $2$, and even parity
in Figs.~\ref{Q410_n2} and \ref{T410_n2}.
For rotating $Q$-balls the scalar field $\f$ must vanish at the
origin, as seen in Eq.~(\ref{erot}).
The energy density $T_{tt}$ of even parity rotating $Q$-balls is torus-like. 
(For odd parity a double torus arises \cite{volkov}.)
The mass $M$, the angular momentum $J$ and the frequency $\omega_s$
of these rotating $Q$-balls 
are exhibited in Table \ref{table2}
together with the properties of the corresponding non-rotating $Q$-ball.
The properties of radially excited rotating $Q$-balls have not yet been
studied \cite{volkov}.
\begin{table}[h!]
\begin{center}
\begin{tabular}{|c|c|c|c|}
\hline
$n$ & $\omega_s$ & $J$ & $M$ \\
\hline
\hline
$0$ & $0.5927$ & $0$ & $293.8$ \\
\hline
$1$ & $0.6931$ & $410$ & $363.4$ \\
\hline
$2$ & $0.7940$ & $820$ & $414.7$ \\
\hline
\end{tabular}
\end{center}
\caption{Physical properties of fundamental $Q$-balls with charge $Q=410$ and
quantum numbers $n=0-2$.}
\label{table2}
\end{table}

\begin{figure}[h!]
\parbox{\textwidth}
{\centerline{
\mbox{
\epsfysize=10.0cm
\includegraphics[width=70mm,angle=0,keepaspectratio]{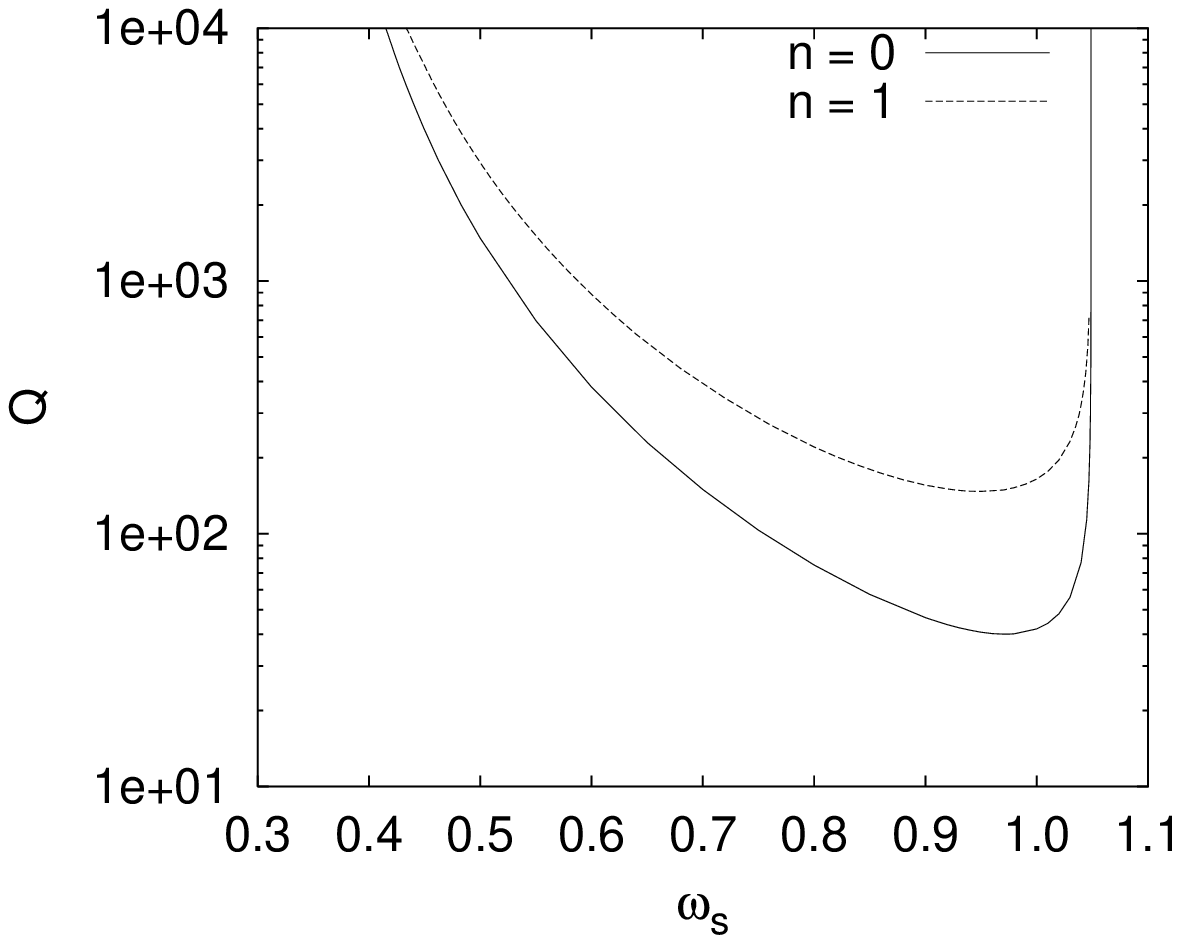}
\includegraphics[width=70mm,angle=0,keepaspectratio]{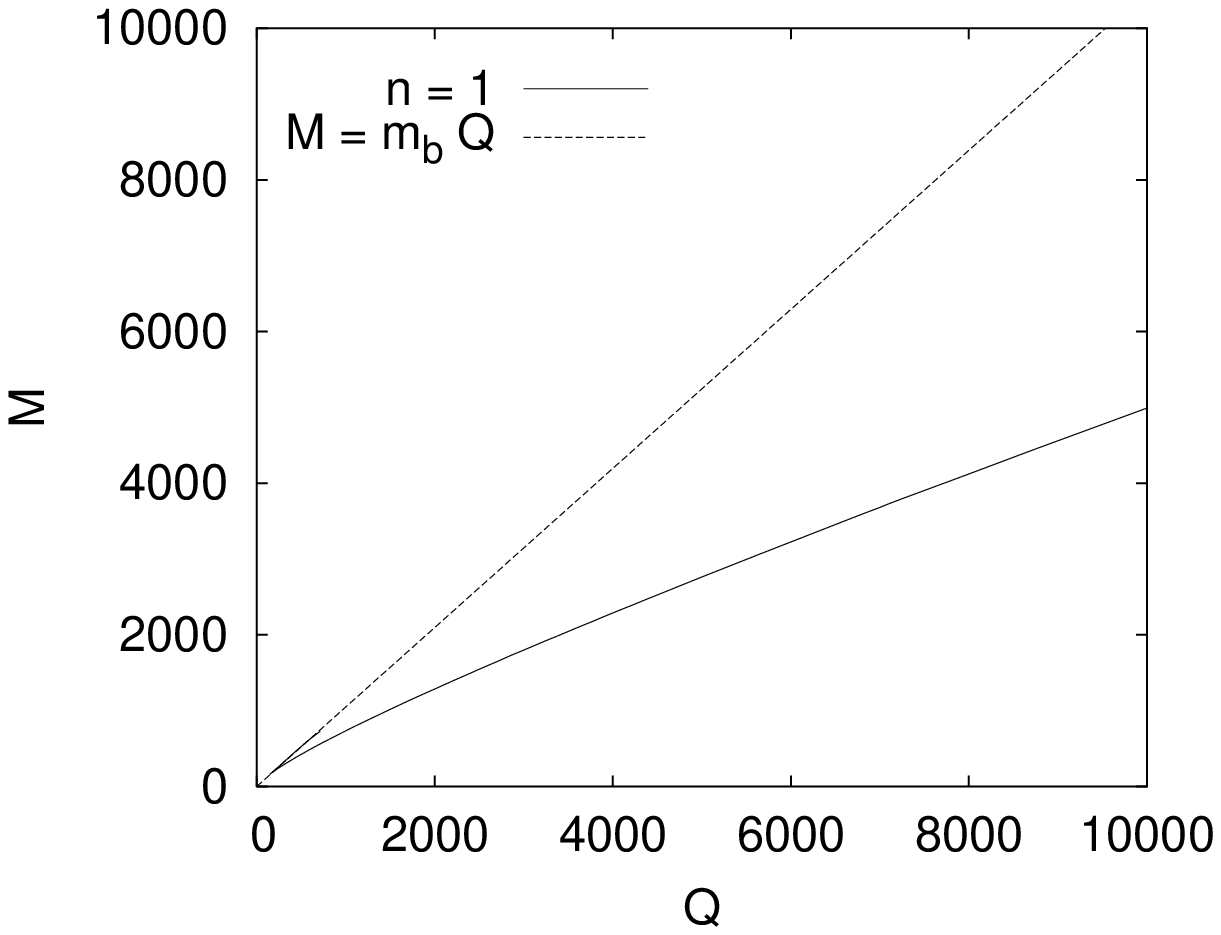}
}}}
\caption{
%12
The charge $Q$ is shown as a function of the frequency $\omega_s$ (left)
and the mass $M$ is shown as a function of the charge $Q$ (right)
for fundamental rotating $Q$-balls with quantum number $n=1$.
For comparison, the charge $Q$ and the mass $M$
also shown for fundamental non-rotating
$Q$-balls. 
The upper branches of the mass $M$ are not discernible (on this scale)
from the mass of $Q$ free bosons, $M=m_{\rm B}Q$ (right).
}
\label{rot-flat}
\end{figure}

We here address the frequency dependence of rotating $Q$-ball solutions.
We focus on fundamental rotating $Q$-balls with quantum number $n=1$ 
and even parity.
In Fig.~\ref{rot-flat} we show the charge $Q$ as a function of the
frequency $\omega_s$ for these fundamental rotating $Q$-balls.
We observe the same upper limiting value $\omega_{\rm max}$, Eq.~(\ref{cond1}),
for the frequency $\omega_s$, as for non-rotating $Q$-balls,
which again ensures asymptotically an exponential fall-off of the
scalar field $\f$.
For a given frequency $\omega_s$ the charge of a rotating
$Q$-ball is larger than the charge of a non-rotating $Q$-ball.
We thus conjecture, that the frequency of rotating $Q$-balls
is also limited by the minimal frequency
$\omega_{\rm min}$, Eq.~(\ref{cond2}).
Unfortunately, for rotating $Q$-balls 
the numerical accuracy decreases considerably 
for large values of the charge $Q$,
refraining us from approaching $\omega_{\rm min}$ more closely.

The mass of the rotating $Q$-balls shows the same cusp structure
as the mass of non-rotating $Q$-balls. 
Only the minimal charge $Q_{\rm min}$
of the rotating $Q$-balls is larger
than the minimal charge of the non-rotating $Q$-balls,
$Q_{\rm min}(n=1) > Q_{\rm min}(n=0)$.
We conclude, that the set of rotating $Q$-balls
exhibits the same general pattern as the set of non-rotating $Q$-balls.

\subsection{Rotating boson stars}

We now turn to rotating boson stars,
obtained when gravity is coupled to the rotating $Q$-ball solutions.
We expect that the set of rotating boson stars also
exhibits the same general pattern as the set of non-rotating boson stars.
Previously \cite{schunck,japan}
rotating boson stars were obtained (only for a $\Phi^4$-potential and)
only for a limited range of the frequency $\omega_s$.
Therefore neither a spiral structure nor cusps were observed.

\begin{figure}[h!]
\parbox{\textwidth}
{\centerline{
\mbox{
\epsfysize=10.0cm
\includegraphics[width=70mm,angle=0,keepaspectratio]{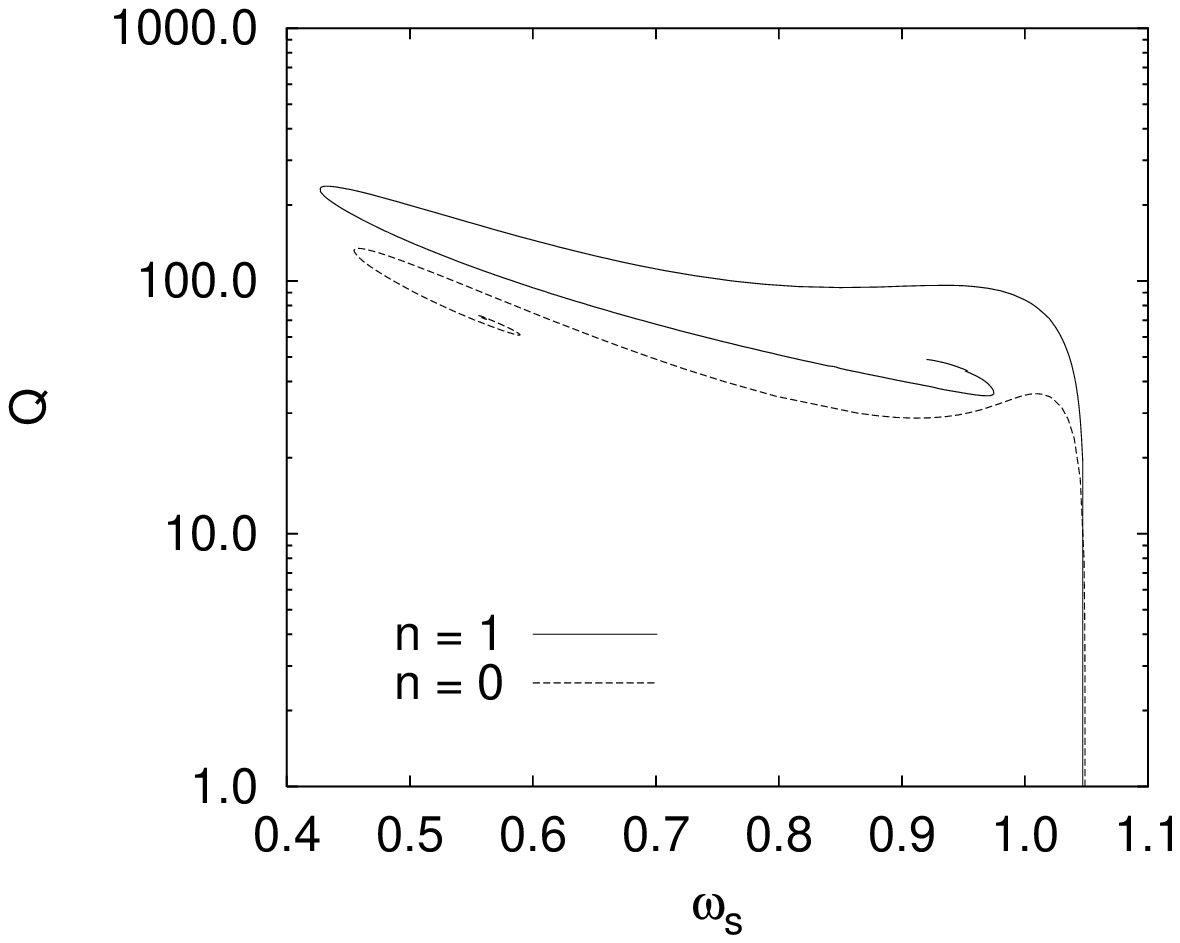}
\includegraphics[width=70mm,angle=0,keepaspectratio]{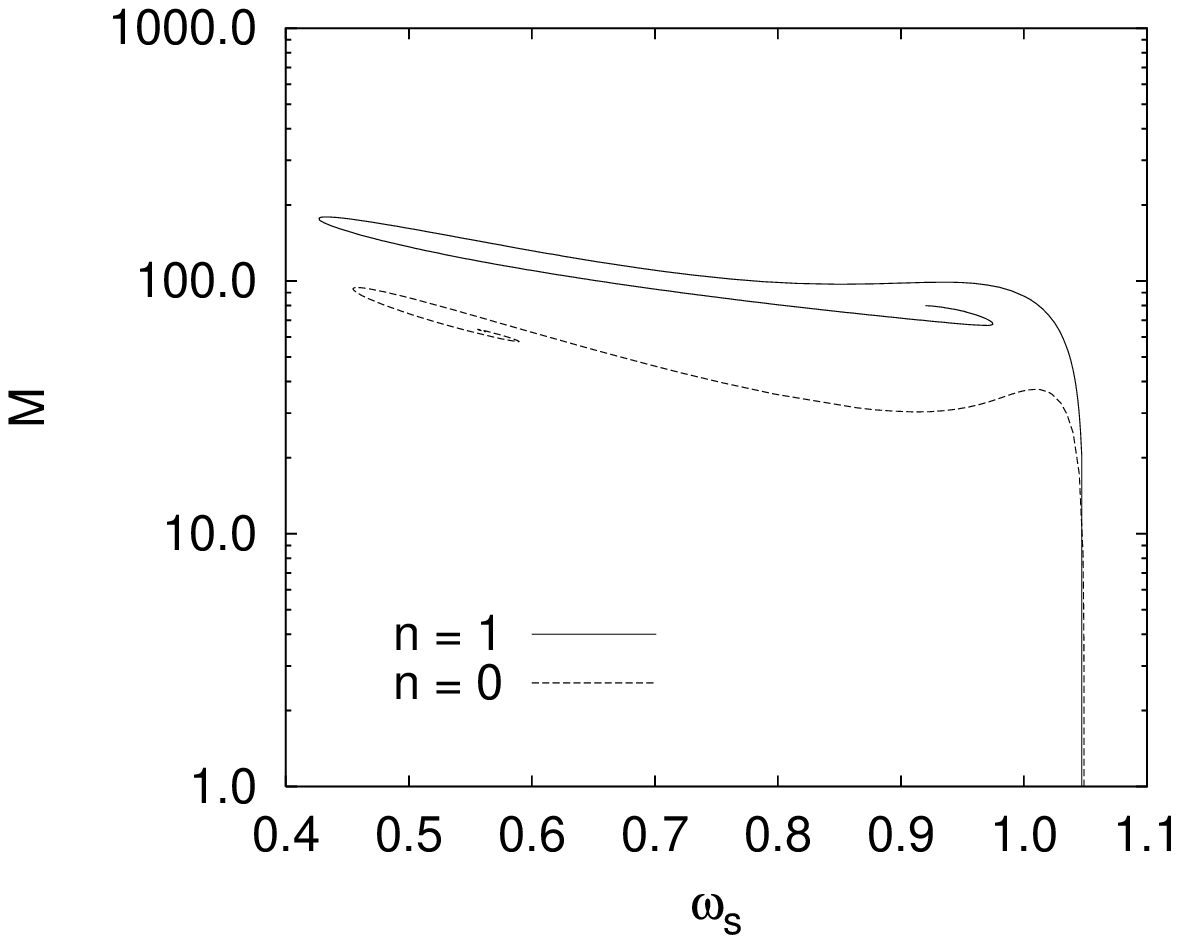}
}}}
\caption{
%13
The charge $Q$ (left) and the mass $M$ (right)
are shown as functions of the frequency $\omega_s$
for fundamental rotating ($n=1$)
and non-rotating boson stars
at the gravitational coupling $\kappa=0.2$.
}
\label{Qrotbs}
\end{figure}

We here focus on fundamental rotating boson stars with rotational
quantum number $n=1$ and even parity.
We exhibit in Fig.~\ref{Qrotbs}
the charge $Q$ as a function of the
frequency $\omega_s$ for fundamental rotating and 
non-rotating boson stars 
at gravitational coupling $\kappa=0.2$.

Again, the frequency $\omega_s$ of the solutions
is bounded from above by
$\omega_{\rm max}$, Eq.~(\ref{cond1}),
ensuring an asymptotically exponential fall-off
of the scalar field.
Furthermore, as for the non-rotating boson stars,
we observe for the rotating boson stars
for the smaller values of the frequency $\omega_s$
a backbending toward larger values of $\omega_s$,
leading apparently also to
an inspiralling of the solutions
towards a limiting solution.
Unfortunately, numerical accuracy does not allow a better
determination of the spiral and the
corresponding limiting values.
Comparison of the rotating and non-rotating sets of solutions shows that
the location of the spiral of the rotating solutions 
is shifted towards larger values of the frequency $\omega_s$.

\begin{figure}[h!]
\parbox{\textwidth}
{\centerline{
\mbox{
\epsfysize=10.0cm
\includegraphics[width=70mm,angle=0,keepaspectratio]{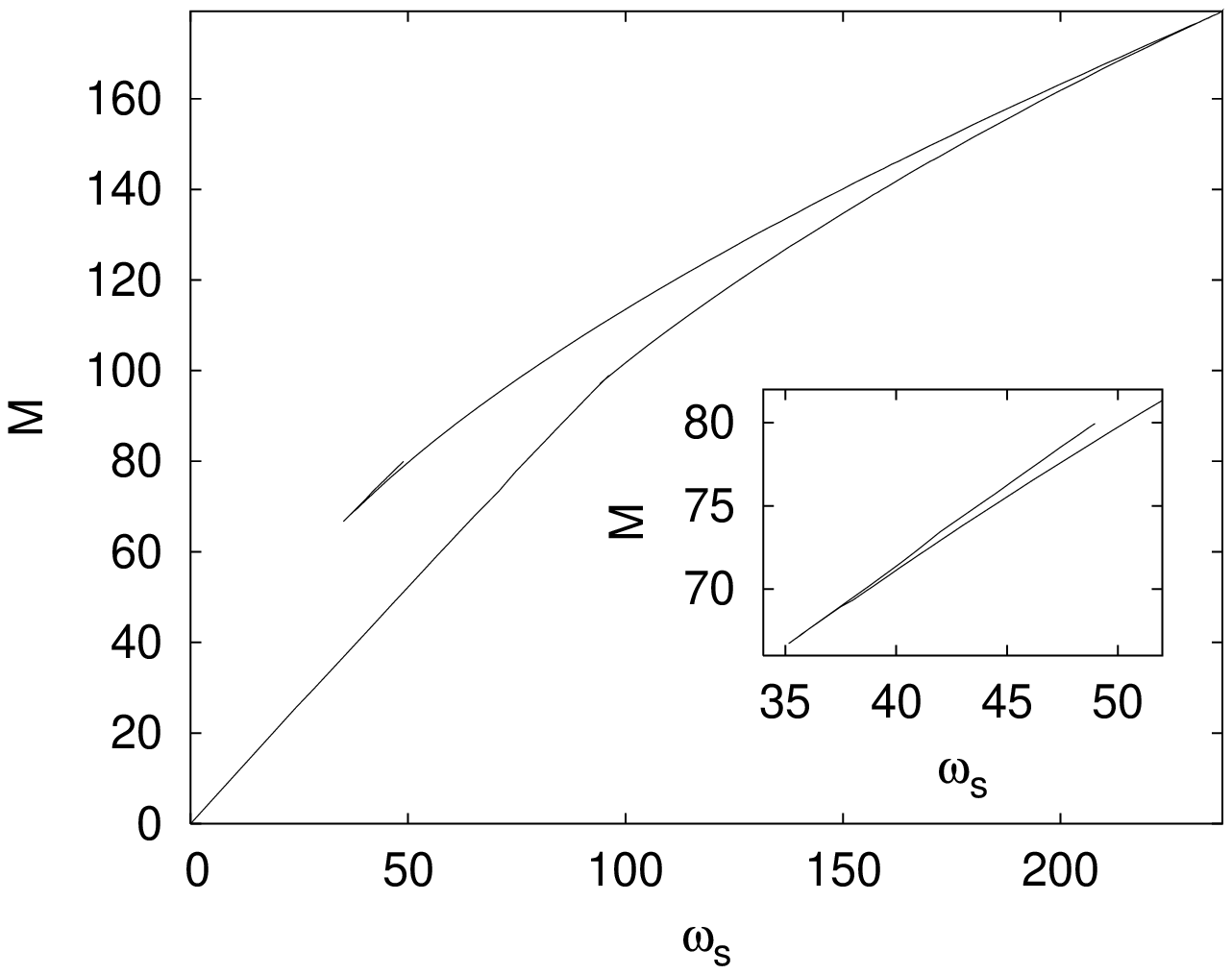}
\includegraphics[width=70mm,angle=0,keepaspectratio]{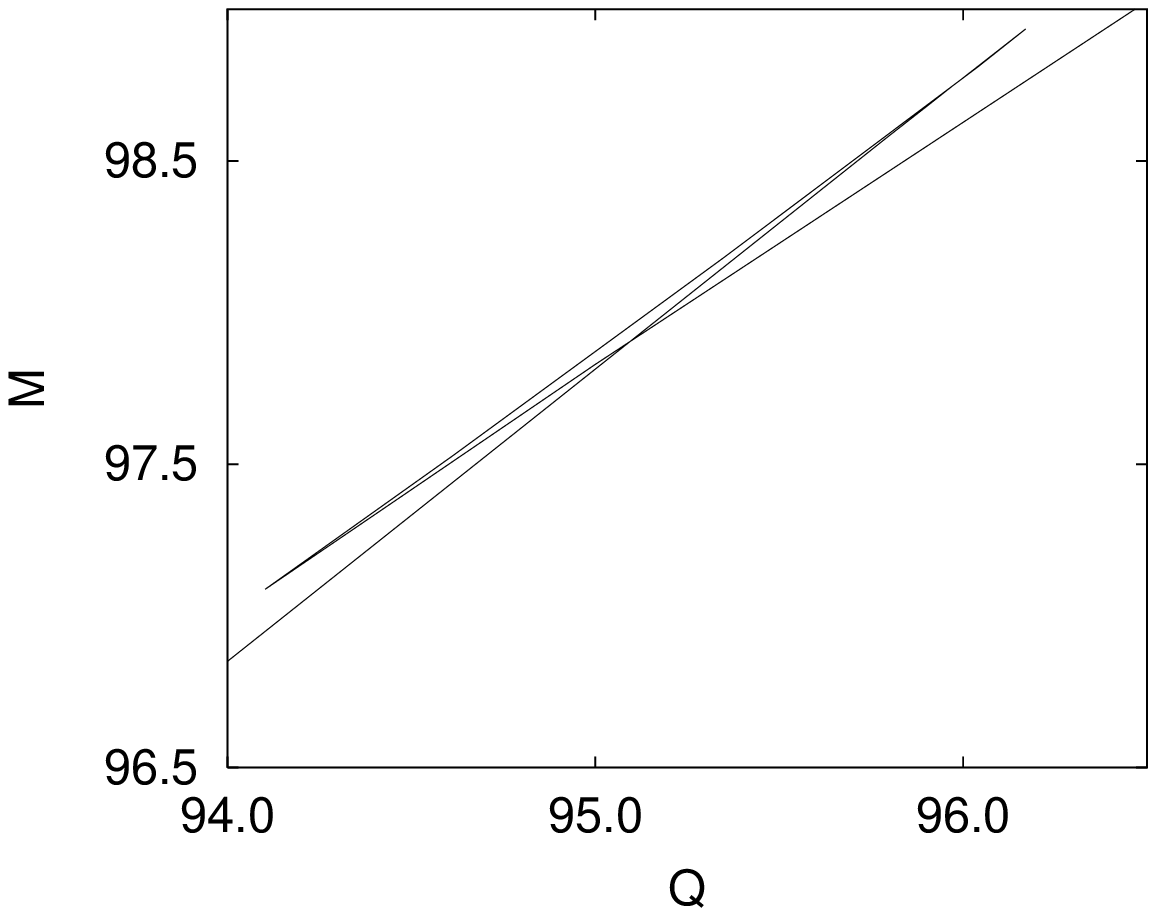}
}}}
\caption{
%14
The mass $M$ 
is shown as a function of the charge $Q$
for fundamental rotating ($n=1$) boson stars
in the full range of existence (left),
and in the charge range of the first set of cusps (right),
for the gravitational coupling constant $\kappa=0.2$.
}
\label{mvsq_rot}
\end{figure}

The mass $M$ has an analogous dependence on the frequency $\omega_s$
as the charge $Q$, as seen in Fig.~\ref{Qrotbs}.
When the mass $M$ of the rotating boson stars
is considered as a function of the charge $Q$,
we observe an analogous cusp structure as for the non-rotating boson stars,
as illustrated in Fig.~\ref{mvsq_rot}.
For smaller values of $\kappa$ (e.g.,~$\kappa=0.2$)
we observe two sets of cusps. The first set of cusps
is again related to the cusp present in flat space, occurring at the
minimal value of the charge $Q_{\rm min}$,
where the upper and lower branch of the rotating $Q$-ball merge.
In curved space the upper branch extends again
only up to a second cusp, reached at 
another critical value of the charge, where this upper branch
merges with a third branch extending down to zero.
For larger values of $\kappa$ (e.g.,~$\kappa=1$)
this set of two cusps, being a
remainder of the flat space solutions, is no longer present.

The second set of cusps is again related to the spirals, 
present only in curved space.
Labelling these cusps again consecutively $N=1,2,...$,
we exhibit the charge $Q$, the mass $M$,
and the frequency $\omega_s$ of the boson star solutions at the first two cusps
in Table~\ref{table_rot_alp},
for the gravitational coupling constants
$\kappa=0.1$, 0.2, and 0.3.
Numerical problems inhibit a better resolution
of this cusp structure (with regard to the larger values of $N$).
\begin{table}[h!]
\begin{center}
\begin{tabular}{|c|c|c|c|c|}
\hline
$N$ & $Q_N$ & $M$ & $J$ & $\omega_s$ \\
\hline
\hline
$1$ & $237.2$ & $179.3$ & $237.3$ & $0.4325$ \\
$2$ & \ $35.2$ & \ $66.7$ & \ $35.1$ & $0.9700$ \\
\hline
\hline
$N$ & $Q_N$ & $M$ & $J$ & $\omega_s$ \\
\hline
\hline
$1$ & $111.0$ & $99.3$ & $111.0$ & $0.5100$ \\
$2$ & \ $23.3$ & $44.4$ & \ $23.3$ & $0.9700$ \\
\hline
\hline
$N$ & $Q_N$ & $M$ & $J$ & $\omega_s$ \\
\hline
\hline
$1$ & $910.9$ & $516.5$ & $911.1$ & $0.3400$ \\
$2$ & \ $76.4$ & $138.8$ & \ $75.7$ & $0.9450$ \\
\hline
\end{tabular}
\end{center}
\caption{Physical characteristics of the first two cusps of the fundamental
rotating ($n=1$) boson star solutions
for gravitational coupling constants $\kappa=0.1$ (upper),
$\kappa=0.2$ (second),
and $\kappa=0.3$ (third).
Deviations from the relation $J=nQ$ are due to numerical
inaccuracy. }
\label{table_rot_alp}
\end{table}

\begin{figure}[h!]
\parbox{\textwidth}
{\centerline{
\mbox{
\epsfysize=10.0cm
\includegraphics[width=70mm,angle=0,keepaspectratio]{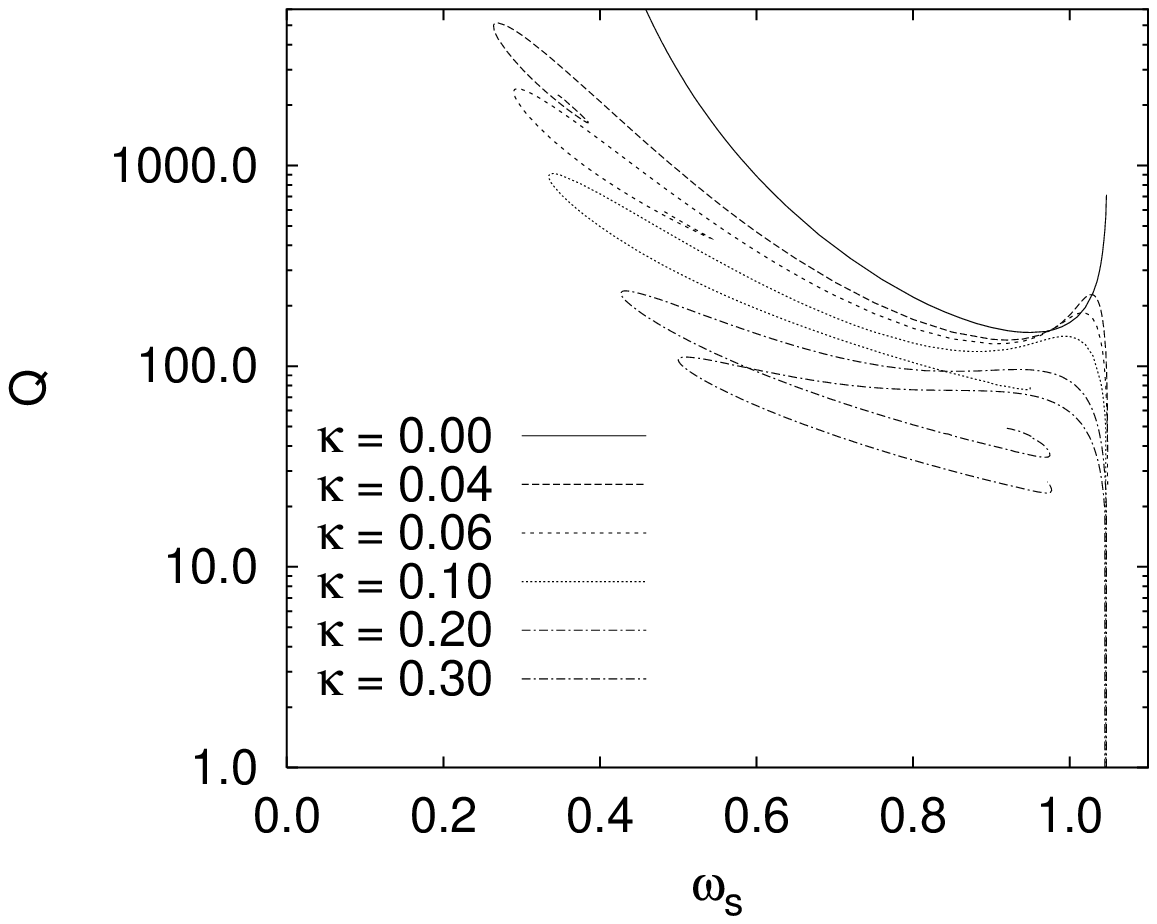}
\includegraphics[width=70mm,angle=0,keepaspectratio]{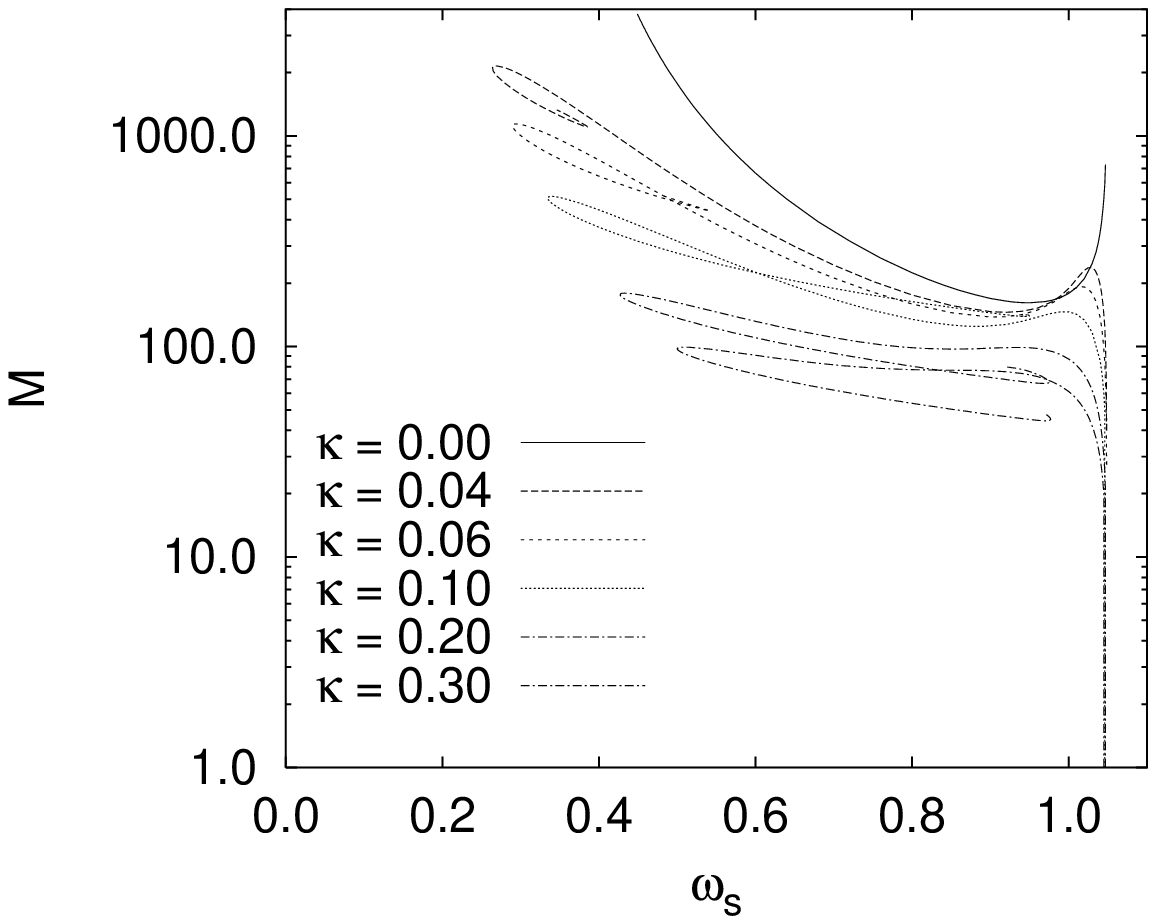}
}}}
\caption{
%15
The charge $Q$ (left) and the mass $M$ (right)
are shown as functions of the frequency $\omega_s$
for the fundamental rotating ($n=1$) boson stars
for several values of the gravitational coupling constant
$\kappa$. Also shown are the limiting flat space values.
}
\label{Qlimrot}
\end{figure}

Addressing finally the $\kappa$-dependence of the domain
of existence of the rotating boson star solutions,
we exhibit in Fig.~\ref{Qlimrot} the charge $Q$
and the mass $M$
as functions of the frequency $\omega_s$ for several values of
the gravitational coupling constant $\kappa$,
including the flat space limit.
Again, we observe that for values of the frequency $\omega_s$ close to
$\omega_{\rm max}$, the flat space values are not approached
monotonically from below with decreasing $\kappa$.

\begin{figure}[h!]
\parbox{\textwidth}
{\centerline{
\mbox{
\epsfysize=10.0cm
\includegraphics[width=70mm,angle=0,keepaspectratio]{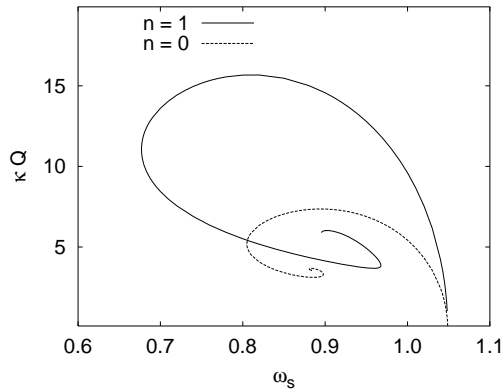}
}}}
\caption{
%15-1
The scaled charge $\kappa Q$ 
is shown as a function of the frequency $\omega_s$
for rotating ($n=1$) boson stars
in the limit $\kappa \rightarrow \infty$ 
For comparison the scaled charge $Q$ is also shown for 
non-rotating boson stars.
}
\label{Qinftyrot}
\end{figure}

We observe from Fig.~\ref{Qlimrot} that also rotating ($n=1$)
boson stars exist only in an interval $[\omega_0(\kappa), \omega_{\rm max}]$.
To obtain the lower bound $\omega_0(\kappa)$ 
in the limit $\kappa \rightarrow \infty$,
we again introduce the scaled
scalar field $\hat{\phi}(r)=\sqrt{\kappa} \phi(r)$,
and substitute $\phi(r)$ in the field equations.
Taking the limit $\kappa \rightarrow \infty$, 
and solving the new set of differential equations,
we find for the rotating ($n=1$) boson stars the lower bound
$\omega_0(\infty) = 0.677$.
The (scaled) charge $\hat{Q} = \kappa Q$
of these solutions is shown in Fig.~\ref{Qinftyrot}.
Concerning the domain of existence in the limit $\kappa \rightarrow 0$, 
we expect a similar pattern as for the non-rotating boson stars.
For $\kappa < 0.06$, for instance, we observe that
$\omega_0(\kappa) < \omega_{\rm min}$.
Numerical problems, however, prevent us from obtaining reliable results
for very small values of $\kappa$.

\section{Conclusions}

We have addressed boson stars and their flat space limit, $Q$-balls.
To illustrate the effects of gravity and rotation,
we have first recalled the main features of non-rotating $Q$-balls.
Second, we have added gravity to obtain non-rotating boson stars.
Third, we have added rotation to obtain rotating $Q$-balls,
and finally we have added both gravity and rotation
to obtain rotating boson stars.
Our main emphasis has been to study the general pattern
displayed by
these regular extended objects, and to determine their domain of existence.

Spherically symmetric $Q$-balls and boson stars
exist only in a limited frequency range.
Whereas both mass and charge of $Q$-balls assume a minimal value
at a critical frequency,
from where they rise monotonically towards both smaller and larger
frequencies,
boson stars show a different type of behaviour.
Their mass and charge tend to zero when the maximal frequency is approached,
while for smaller values of the frequency
charge and mass exhibit a spiral-like frequency dependence,
leading to limiting solutions with finite values of 
the mass $M_{\rm lim}$, the charge $Q_{\rm lim}$ 
and the frequency $\omega_{\rm lim}$, 
depending on the gravitational coupling constant $\kappa$.

Denoting the domain of existence of 
spherically symmetric boson stars by
$[\omega_0(\kappa), \omega_{\rm max}]$,
we observe, that when $\kappa \rightarrow \infty$,
the lower limiting value of the frequency $\omega_0(\kappa)$ 
tends to a finite value $\omega_0(\infty) < \omega_{\rm max}$,
obtained by solving the set of field equations for the scaled scalar field
$\sqrt{\kappa} \phi(r)$.

For small values of $\kappa$, on the other hand,
the domain of existence consists of two parts,
$[\omega_0(\kappa), \omega_{\rm min}]$ and
$[\omega_{\rm min}, \omega_{\rm max}]$.
In the limit $\kappa \rightarrow 0$, 
the lower limiting value of the frequency $\omega_0(\kappa)$
tends to a finite value, $\omega_0(0) < \omega_{\rm min}$.
In the interval $[\omega_0(0), \omega_{\rm min}]$
there are no flat space solutions,
possessing finite charge or mass.
The limiting values of the spirals,
$Q_{\rm lim}$ and $M_{\rm lim}$ diverge with $\kappa^{-3/2}$
when $\kappa \rightarrow 0$.
(For a potential with degenerate minima \cite{lee-bs}
only the single interval
$[\omega_{\rm min}, \omega_{\rm max}]$ needs to be considered,
since $\omega_{\rm min}=0$.)

Spherically symmetric $Q$-balls and boson stars
possess radial excitations. Their systematic study has not yet been
performed. We have, however, considered 
the first radial excitations of $Q$-balls and the first radial excitations of
boson stars at a given gravitational coupling.
This indicates, that radially excited $Q$-balls and boson stars
possess an analogous frequency dependence and critical behaviour
as the corresponding fundamental solutions.

Rotating $Q$-balls and boson stars possess a quantized angular momentum,
$J=nQ$. For each rotational quantum number $n$ there are
even and odd parity solutions.
Here we have focussed on solutions with $n=1$ and even parity.
These rotating $Q$-balls show a frequency dependence
and critical behaviour analogous to the non-rotating $Q$-balls.
Likewise, the rotating boson stars show a frequency dependence
and critical behaviour analogous to the non-rotating boson stars.
In particular, on the one hand we observe that 
the mass and charge of rotating boson stars
tend to zero when the maximal frequency is approached,
and on the other hand we see
a spiral-like frequency dependence for the mass and charge
for the smaller values of the frequency.
Unfortunately, numerical inaccuracies increase along the spirals,
preventing a reliable construction of their interior parts.

Rotating ($n=1$) boson stars also exist only 
in an interval $[\omega_0(\kappa), \omega_{\rm max}]$.
In the range of $\kappa$ considered,
the lower limiting value $\omega_0(\kappa)$ of
rotating boson star solutions
is larger than 
the lower limiting value $\omega_0(\kappa)$ of
non-rotating boson star solutions.
In the limit $\kappa \rightarrow \infty$,
the lower bound $\omega_0(\kappa)$ 
is again obtained by solving the set of field equations 
for the scaled scalar field $\sqrt{\kappa} \phi(r)$.

For small values of $\kappa$, 
we also observe an analogous pattern
for the rotating boson stars as for the non-rotating boson stars.
Numerical problems prevent us, however, from obtaining reliable results
for the domain of existence for very small values of $\kappa$.

A systematic study of the frequency dependence of 
rotating boson stars with higher rotational quantum number $n$
and with both parities is still missing
and represents a numerical challenge.
While the energy density of rotating boson stars with
$n=1$ and even parity has a torus-like structure,
the energy density of rotating boson stars with $n=1$ and odd
parity should have a double-torus structure.
For radially excited rotating boson stars the structure 
of the energy density may involve concentric tori.
Their construction remains an open challenge, as well.

\begin{acknowledgments}
B.K. gratefully acknowledges support by the DFG under contract
KU612/9-1.
\end{acknowledgments}

\begin{appendix}

\section{Auxiliary Functions}

In order to be able to calculate boson star solutions
for a given value of the charge $Q$,
we define 2 auxiliary functions, $\rho(r, \t)$ and $\omega_s(r, \t)$,
The equation of motion for the function $\rho(r, \t)$
consists of the Laplacian acting on $\rho(r, \t)$ and a source term, 
which is proportional to the charge density, Eq.~(\ref{Qc}),
\begin{equation}
\left(|g|^{1/2}g^{rr}\rho_{, \, r} \right)_{,r}+\left(|g|^{1/2}
g^{\t \t}\rho_{, \, \t} \right)_{, \, \t} = -|g|^{1/2}\left(g^{tt}+g^{t \varphi}\frac{n}{\omega_s}\right) \f ^2  
\ . \end{equation}
Integration yields
\begin{multline}
\int _0 ^{\pi}\left(|g|^{1/2}g^{rr}\rho_{, \, r} \right) \, \bigg|_0^{\infty} \, d\t+ \int _0^{\infty}\left(|g|^{1/2}
g^{\t\t}\rho_{, \, \t} \right) \, \bigg|_0^{\pi} \, dr \\
= -\int_0^{\infty}\int_0^{\pi}
|g|^{1/2} \left(g^{tt}+g^{t \varphi}\frac{n}{\omega_s}\right) 
 \f^2 \, dr \, d\t  \label{1}
\ , \end{multline}
and with $|g|^{1/2}=l^{3/2}gr^2\sin{\t}/f$ and Eq.~(\ref{Qc})
we obtain
\begin{eqnarray}
\int _0^{\pi} \left( \sin{\t} \, l^{1/2}\, r^2 \rho_{,r} \right) \, \bigg|_0^{\infty} \, d \t + \int _0 ^{\infty} \left( \sin{\t} \,
l^{1/2}\,  \rho_{,\t} \right) \, \bigg|_0^{\pi} \, dr 
&=&\frac{Q}{4 \pi \omega_s} \ . \label{integral4}
\end{eqnarray}
Making use of the asymptotic expansion for the function $\rho(r, \t)$,
\begin{eqnarray}
\rho = \rho_{\infty} + \frac{C}{r} + O\left(\frac{1}{r^2}\right)
 + \dots \ ,
\end{eqnarray}
with constants $C$ and $\rho_{\infty}$, and thus
asymptotically $r^2\rho_{, \, r} \longrightarrow -C$,
we obtain 
\begin{eqnarray}
C\int_0^{\pi} \sin{\t} \, d\t &=&
 2C = \frac{Q}{4 \pi \omega_s}
\ . \end{eqnarray}
This then yields the connection between the function $\rho(r,\t)$ at infinity,
the frequency $\omega_s$ and the charge $Q$,
\begin{equation}
\left( r^2 \rho _{, \, r} \omega_s-\frac{Q}{8\pi} \right) 
\bigg|_{r \rightarrow \infty}= 0 \ .
\label{rhoomQ} \end{equation} 
Since we cannot impose the value of the frequency $\omega_s$, 
when we impose the value of the charge, we need to solve
for the frequency $\omega_s$. The frequency $\omega_s$
is a constant. An adequate equation is therefore,
that the Laplacian of $\omega_s$ vanishes,
\begin{eqnarray}
\left( |g| ^{1/2} \, g^{rr} \, \omega_{s,\, r} \right) _{, \, r} + \left(|g| ^{1/2} \, g^{\t\t} \, \omega_{s, \, \t} \right)_{, \, \t} &=& 0
\ . \label{dgl_omegas}
\end{eqnarray}

The appropriate set of boundary conditions for these two
auxiliary functions are
\begin{equation}
\partial_r \omega_s|_{r=0}=0 \ , \ \ \
8\pi r^2 \rho _{, \, r} \omega_s|_{r \rightarrow \infty}=Q \ , \ \ \
\partial_{\t} \omega_s|_{\t=0}=0 \ , \ \ \
\partial_{\t} \omega_s|_{\t=\pi/2}=0 \ ,
\end{equation}
\begin{equation}
\partial_r \rho|_{r=0}=0 \ , \ \ \
 \rho|_{r \rightarrow \infty}=\rho_\infty \ , \ \ \
\partial_{\t} \rho|_{\t=0}=0 \ , \ \ \
\partial_{\t} \rho|_{\t=\pi/2}=0 \ ,
\end{equation}
and $\rho_\infty$ is an arbitrary constant.

When instead of the charge $Q$ a value for the frequency $\omega_s$
is imposed, use of the auxiliary functions is not necessary,
but convenient, since the charge is then obtained directly.
In this case, the same set of boundary conditions is appropriate
except for 
\begin{equation}
\omega_s|_{r \rightarrow \infty}= \omega_\infty \ ,
\end{equation}
and $\omega_\infty$ is the required value.

\section{System of differential equations}

\subsection{Spherically symmetric solutions}\label{dgl_sys1}

For the spherically symmetric boson stars
we obtain the following set of coupled non-linear 
ordinary differential equations 
\begin{eqnarray}
%Gleichung für f_rr
\partial_r^2 f &=& - \frac{1}{2} \, \frac{1}{f \, l} \, \biggl[ 4 \kappa f l^2
\, U(\f) - 8 \kappa \omega_s^2 \, l^2 \, \f^2 +\frac{4}{r} \, f l \partial_r
f - 2 l \partial_r f \nonumber \\
&&\phantom{- \frac{1}{2} \, \frac{1}{f \, l} \, \biggl[} + f \partial_r l \, \partial_r f \biggr] \label{f_rr} 
%Gleichung für l_rr
\end{eqnarray}
\begin{eqnarray}
\partial_r^2 l &=& - \frac{1}{2} \, \frac{1}{f \, l} \, \biggl[ 8 \kappa l^3 \, U(\f) - 8 \kappa \omega_s^2 \, 
\frac{l^3}{f} \, \,\f^2 + \frac{6}{r} \, f l
\partial_r l \nonumber \\
&&\phantom{-\frac{1}{2} \, \frac{1}{f \, l} \, \biggl[} - f \left( \partial_r l \right)^2 \biggr] \label{l_rr} 
%Gleichung für phi_rr
\end{eqnarray}
\begin{eqnarray}
\partial_r^2\f &=& -\frac{1}{2} \, \frac{1}{f \, l} \, \biggl[ 2 \omega_s^2 \,
\frac{l^2}{f} \, \f -l^2 \, \frac{\partial U(\f)}{\partial \f} +
\frac{4}{r} \, f l \partial_r \f \nonumber \\
&&\phantom{-\frac{1}{2} \, \frac{1}{f \, l} \, \biggl[} +f \partial_r l \, \partial_r \f \biggr] \label{phi_rr} \ .
\end{eqnarray}

For the auxiliary functions the equations are
\begin{eqnarray}
0&=& - \frac{f}{l} \left( \partial_r^2 \rho + \frac{2}{r} \, \partial_r \rho + \frac{1}{2\, l} \, \partial _r l \,
\partial_r \rho \right) + \frac{1}{f} \f^2 \quad , \label{rho_ohnerot}
\end{eqnarray} 
\begin{eqnarray}
0&=& \frac{f}{l} \left( \partial_r^2 \omega_s + \frac{2}{r} \, 
 \partial _r \omega_s + \frac{1}{2\, l} \, \partial _r l  
\, \partial_r \omega_s \right) \quad . \label{omegas_ohnerot}
\end{eqnarray}

\subsection{Axially symmetric solutions}\label{dgl_sys2}

For the rotating boson stars
we obtain the following set of coupled non-linear 
partial differential equations 
\begin{eqnarray}
%Gleichung für f
r^2 \, \partial_r^2 f + \partial_{\t}^2 f &=& - \frac{1}{2} \, \frac{1}{f \, l}
\, \biggl[ 4 \kappa r^2 \, f l^2 \, U(\f) - 8 \kappa n^2 \, l^2 \, g \omega^2 \,
\f^2 -16 \kappa r n \omega_s l^2 \, g \omega \f^2 \nonumber \\
&&\phantom{- \frac{1}{2} \, \frac{1}{f \, l}
\, \biggl[} -8 \kappa r^2 \omega_s^2 \,
l^2 \, g \f^2 - 2 \sin^2 \t \,  \, l^2 \, \omega^2 + 4 r f l \partial_r f 
\nonumber \\
&&\phantom{- \frac{1}{2} \, \frac{1}{f \, l}
\, \biggl[} + 2 \frac{\cos{\t}}{\sin{\t}} \, f l
\partial_{\t} f + fl \partial_{\t} f - 2 l \left( r^2 \, \left( \partial_r f
\right)^2 \, + \left( \partial_{\t} f \right)^2 \right) \nonumber \\
&&\phantom{- \frac{1}{2} \, \frac{1}{f \, l}
\, \biggl[} +  r^2 \, f \partial_r l
\, \partial_r f + 4 r \sin^2 \t \,  \, l^2 \, \omega \partial_r \omega \nonumber
\\
&&\phantom{- \frac{1}{2} \, \frac{1}{f \, l}
\, \biggl[} - 2 \sin^2 \t \,  \, l^2 \, \left( r^2 \,
\left( \partial_r \omega \right)^2 \, + \left( \partial_{\t} \omega \right)^2
\right) \biggr] \label{dgl_rtf} 
\end{eqnarray}
\begin{eqnarray}
%Gleichung für g
r^2 \, \partial_r^2 g + \partial_{\t}^2 g &=& - \frac{1}{2} \, \frac{1}{f \, l}
\, \biggl[ -8 \kappa n^2 \, f l g^2 \, \frac{1}{\sin^{2} \t}  \, \f^2 - 2 \sin^2 \t \,  \,
\frac{l^2}{f} \, g \omega^2 \nonumber \\
&&\phantom{- \frac{1}{2} \, \frac{1}{f \, l}
\, \biggl[} + 8 \kappa f l g \left( \partial_{\t} \f
\right)^2+ 4r f l \partial_r g - 2 \frac{\cos{\t}}{\sin{\t}} \, f l
\partial_{\t} g \nonumber \\
&&\phantom{- \frac{1}{2} \, \frac{1}{f \, l}
\, \biggl[} -2 \frac{f l}{ g} \, \left( r^2 \, \left( \partial_r g \right) ^2 + \left(
\partial_{\t} g \right)^2 \right) + f \left( r^2 \, \partial_r l \, \partial_r g
- \partial_{\t} l \, \partial_{\t} g \right) \nonumber \\
&&\phantom{- \frac{1}{2} \, \frac{1}{f \, l}
\, \biggl[} + 2 \frac{l g}{f} \left(
\partial_{\t} f \right)^2 - 2 \frac{\cos{\t}}{\sin{\t}} \, f
g \partial_{\t} l - 3 \frac{f}{l} \, g \left( \partial_{\t} l \right) ^2 + 2 f g
\partial_{\t}^2 l \nonumber \\
&&\phantom{- \frac{1}{2} \, \frac{1}{f \, l}
\, \biggl[} + 4 r \sin^2 \t \,  \, \frac{l^2g}{f} \,  \omega \partial_r
\omega \nonumber \\
&&\phantom{- \frac{1}{2} \, \frac{1}{f \, l}
\, \biggl[} -2 \sin^2 \t \,  \, \frac{l^2g}{f} \, \left( r^2 \, \left( \partial_r
\omega \right)^2 \, + 2 \left( \partial_{\t} \omega \right)^2 \right) \biggr] \label{dgl_rtg} 
\end{eqnarray}
\begin{eqnarray}
%Gleichung für l
r^2 \, \partial_r^2 l + \partial_{\t}^2 l &=& - \frac{1}{2} \, \frac{1}{f \, l }
\, \biggl[ 8 \kappa r^2 \, l^3 \, g U(\f) + 8 \kappa \frac{1}{\sin^{2} \t}  \, n^2 \, f
l^2 \, g \f^2 - 8 \kappa n^2 \, \frac{l^3g}{f} \,  \omega^2 \, \f^2 \nonumber
\\
&&\phantom{- \frac{1}{2} \, \frac{1}{f \, l}
\, \biggl[} - 16 \kappa
r n \omega_s \frac{l^3g}{f} \,  \omega \f^2 - 8 \kappa r^2 \, \omega_s^2 
\frac{l^3g}{f} \,  \f^2 + 6 r f l \partial_r l \nonumber \\
&&\phantom{- \frac{1}{2} \, \frac{1}{f \, l}
\, \biggl[} + 4 \frac{\cos{\t}}{\sin{\t}}\, f l
\partial_{\t} l - f \left( r^2 \, \left( \partial_r l \right)^2 + \left(
\partial_{\t} l \right)^2 \right) \biggr] \label{dgl_rtl} 
\end{eqnarray}
\begin{eqnarray}
%Gleichung für omega
r^2 \, \partial_r^2 \omega + \partial_{\t}^2 \omega &=& - \frac{1}{2} \,
\frac{1}{f \, l} \, \biggl[ - 8 \kappa \frac{1}{\sin^{2} \t}
  \, n^2  \, f l g \omega
\f^2 - 8 \kappa r \frac{1}{\sin^{2} \t}  \, n \omega_s f l g \f^2 \nonumber \\
&&\phantom{- \frac{1}{2} \, \frac{1}{f \, l}
\, \biggl[} - 4 f l \omega + 4 r f
l \partial_r \omega + 6 \frac{\cos{\t}}{\sin{\t}} \, f l \partial_{\t} \omega \nonumber \\
&&\phantom{- \frac{1}{2} \, \frac{1}{f \, l}
\, \biggl[} - 4
l \left( r^2 \, \partial_r f \, \partial_r \omega + \partial_{\t} f \,
\partial_{\t} \omega \right) \nonumber \\
&&\phantom{- \frac{1}{2} \, \frac{1}{f \, l}
\, \biggl[} + 3 f \left( r^2 \, \partial_r l \, \partial_r
\omega + \partial_{\t} l \, \partial_{\t} \omega \right) \nonumber \\
&&\phantom{- \frac{1}{2} \, \frac{1}{f \, l}
\, \biggl[} + 4 r l \omega
\partial_r f - 3 r f \omega \partial_r l \biggr] \label{dgl_rto} 
\end{eqnarray}
\begin{eqnarray}
%Gleichung für phi
r^2 \, \partial_r^2 \f + \partial_{\t}^2 \f &=& - \frac{1}{2} \, \frac{1}{f \, l}
\, \biggl[ - r^2 \, l^2 \, g \frac{\partial U(\f)}{\partial \f} -2 
\frac{1}{\sin^{2} \t}
\, n^2 \, f l g \f \nonumber \\
&&\phantom{- \frac{1}{2} \, \frac{1}{f \, l} \, \biggl[} + 2 \frac{l^2g}{f} \,  \left( \omega n + r \omega_s \right)^2 \f + 4 r f l
\partial_r \f \nonumber \\
&&\phantom{- \frac{1}{2} \, \frac{1}{f \, l}
\, \biggl[} + 2\frac{\cos{\t}}{\sin{\t}} \, f l
\partial_{\t} \f + f \left( r^2 \, \partial_r l \, \partial_r \f + \partial_{\t}
l \, \partial_{\t} \f \right) \biggr] \label{dgl_rtphi} \ .
\end{eqnarray}

For the auxiliary functions the equations are
\begin{eqnarray}
0&=& - \phantom{\frac{1}{r^2} \,} \frac{f}{l \, g} \left(\, \partial_r ^2 \rho + \frac{2}{r} \, \partial_r \rho + \frac{1}{2\, l} \, \partial_r l \, \partial_r
\rho \right) \nonumber \\
&& - \frac{1}{r^2} \, \frac{f}{l \, g} \, \left( \, \partial _{\t} ^2 \rho + \frac{\cos{\t}}{\sin{\t}} \, \partial _{\t} \rho + \frac{1}{2\, l} \,
\partial_{\t} l \, \partial _{\t} \rho \right) \nonumber \\
&& + \frac{1}{f} \left( 1 + \frac{1}{r} \frac{\omega}{\omega_s} n \right) \f^2 \quad , \label{rho_mitrot}
\end{eqnarray}
\begin{eqnarray}
0&=& \frac{f}{l \, g} \left( \partial_r^2 \omega_s + \frac{2}{r} \, \partial _r \omega_s + \frac{1}{2\, l} \, \partial
_r l  \, \partial_r \omega_s \right) \nonumber \\
&& + \frac{1}{r^2} \, \frac{f}{l \, g} \left( \partial_{\t}^2 \omega_s + \frac{\cos{\t}}{\sin{\t}}\, \partial_{\t} \omega_s + \frac{1}{2\, l} \,
\partial_{\t} l \, \partial_{\t} \omega_s \right) \quad . \label{omegas_mitrot}
\end{eqnarray}

\section{Stress-energy-tensor}

\subsection{Spherically symmetric solutions}\label{ei}

The nonvanishing components of the stress-energy--tensor for non-rotating
boson stars read
\begin{eqnarray}
T_{tt}&=&\left( f U(\f)+\omega_s^2 \, \f^2 + \frac{f^2 }{l} \, \left( \partial_r \f \right) ^2 \right) \\ 
T_{rr}&=& -\left( \frac{l}{f} U(\f) - \frac{l}{f^2} \omega_s^2 \, \f^2 - \left( \partial _r \f \right)^2 \right) \\
T_{\t\t} &= & - r^2 \left(\frac{l}{f} U(\f) - \frac{l}{f^2}  \omega_s^2 \f^2 + \left( \partial_r \f \right) ^2 \right) \\
T_{\varphi \varphi} &=& - r^2 \, \sin^2 \t \,  \, l \left( \frac{1}{f} \, U(\f) - \frac{1}{f^2} \, \omega^2 \, \f ^2 + \frac{1}{l} \, \left( \partial _r \f \right) ^2 \right) 
\end{eqnarray}

\subsection{Axially symmetric solutions}\label{ei_rot}

The nonvanishing components of the stress-energy--tensor for rotating
boson stars read
\begin{eqnarray}
T_{tt} & = & \biggl[ f \, U(\f) - \sin^2 \t \, \frac{l}{f} \omega^2 \, U(\f )\nonumber \\
&& \phantom{\biggl[ }+ \omega_s^2 \, \f^2 -\frac{2}{r} \, \omega n \omega_s \, \f^2 + \frac{1}{r^2 \, \sin^2 \t \, } \, \frac{f^2}{l} \, n^2 \, \f^2 \nonumber \\
&& \nonumber \\
&&\phantom{\biggl[ } + \frac{2}{r} \, \sin^2 \t \, \frac{l}{f^2} \omega^3 \, n \omega_s \f^2 + \sin^2 \t \, \frac{l}{f^2} \omega^2 \, \omega_s^2 \, \f^2 \nonumber \\ 
&& \nonumber \\
&& \phantom{\biggl[ }- \frac{2}{r^2} \, \omega^2 \, n^2 \, \f^2 + \frac{1}{r^2} \, \sin^2 \t \, \frac{l}{f^2} \omega^4 \, n^2 \, \f ^2 \nonumber \\
&& \nonumber \\
&& \phantom{\biggl[ }+ \frac{f^2}{l\, g} \, \left( \partial_r \f \right) ^2 - \sin^2 \t \, \frac{1}{g} \, \omega^2 \, \left( \partial_r \f \right) ^2 \nonumber \\
&& \nonumber \\
&& \phantom{\biggl[ }+ \frac{1}{r^2} \, \frac{f^2 }{l \, g} \, \left( \partial_{\t} \f \right)^2 - \frac{1}{r^2} \, \sin^2 \t \,  \, \frac{l}{g}
\omega^2 \left( \partial_{\t} \f \right) ^2  \biggr] 
\end{eqnarray}
\begin{eqnarray}
T_{rr} &=&\biggl[ - \frac{lg}{f} U(\f) + \frac{lg}{f^2}  \omega_s^2 \, \f^2 + \frac{2}{r} \, \frac{lg}{f^2}  \omega n \omega_s \f^2 \nonumber \\
&& \phantom{\biggl[ }- \frac{1}{r^2 \, \sin^2 \t \,  } \, g n^2 \, \f ^2 + \frac{1}{r^2 } \, \frac{lg}{f^2}  \omega^2 \, n^2 \, \f^2 \nonumber \\
&& \nonumber \\
&&\phantom{\biggl[ } + \left( \partial _r \f \right) ^2 - \frac{1}{r^2 } \, \left( \partial _{\t} \f \right) ^2 \biggr] 
\end{eqnarray}
\begin{eqnarray}
T_{r\t} &=&\bigg[2 \left( \partial_{r} \f \right) \, \left( \partial _{\t} \f \right) \biggr] 
\end{eqnarray}
\begin{eqnarray}
T_{\t\t}& =&\biggl[ - r^2 \, \frac{lg}{f} U(\f) + r^2 \, \frac{lg}{f^2}  \omega_s^2 \, \f ^2 \nonumber \\
%&& \nonumber \\
&& \phantom{\biggl[ }+ 2r \, \frac{lg}{f^2}  \omega n \omega_s \f^2 + \frac{l g}{f^2}  \omega^2 \, n^2 \, \f^2  \nonumber \\
&& \nonumber \\
&&\phantom{\biggl[ }- \frac{1}{\sin^2 \t \, } \, g n^2 \, \f^2 - r^2 \left( \partial_r \f \right) ^2 + \left( \partial_{\t} \f \right) ^2
\biggr] 
\end{eqnarray}
\begin{eqnarray}
T_{t \varphi} &=& \biggl[  r \sin^2 \t \, \frac{l}{f} \omega U(\f) - r \sin^2 \t \, \frac{l}{f^2} \omega \omega_s^2 \, \f^2 \nonumber \\
%&& \nonumber \\
&&\phantom{\biggl[ } -2 \sin^2 \t \,  \, \frac{l}{f^2} \omega^2 \, n \omega_s \f^2 + \frac{1}{r} \, \omega n^2 \, \f^2 \nonumber \\
&& \nonumber \\
&&\phantom{\biggl[ } - \frac{1}{r} \, \sin^2 \t \, \frac{l}{f^2} \omega^3 \, n^2 \, \f ^2 + 2 n \omega_s \f^2 \nonumber \\
&& \nonumber \\
&& \phantom{\biggl[ }+ r \sin^2 \t \,  \, \frac{1}{g} \, \omega \left( \partial_r \f \right) ^2 + \frac{1}{r} \, \sin^2 \t \,  \, \frac{1}{g} \, \omega \left( \partial_{\t} \f \right) ^2 \biggr] 
\end{eqnarray}
\begin{eqnarray}
T_{\varphi \varphi}&=& \biggl[  -r^2 \, \sin^2 \t \,  \, \frac{l}{f} U(\f)+ r^2 \, \sin^2 \t \,  \, \frac{l}{f^2} \omega_s^2 \, \f ^2 \nonumber \\
%&& \nonumber \\
&& \phantom{\biggl[ }+ 2r \sin^2 \t \,  \, \frac{l}{f^2} \omega n \omega_s \f ^2 + n^2 \, \f^2 \nonumber \\
&& \nonumber \\
&& \phantom{\biggl[ }+ \sin^2 \t \,  \, \frac{l}{f^2} \omega^2 \, n^2 \, \f^2 \nonumber \\
&& \nonumber \\
&& \phantom{\biggl[ }- r^2 \, \sin^2 \t \,  \, \frac{1}{g} \, \left( \partial_r \f \right) ^2 - \sin^2 \t \,  \, \frac{1}{g} \, \left( \partial
_{\t} \f \right) ^2 \biggr] \ .
\end{eqnarray}

\end{appendix}
%\clearpage


\begin{thebibliography}{99}
\bibitem{foot1}
In all formulae of \cite{volkov} based on the scalar field equation
a factor of two is missing in front of the potential $U$.
\bibitem{foot2}
The numerical solutions shown in \cite{volkov} can be considered as
obtained with a potential differing from the true potential
by a factor of two.
The expression for the mass employed in \cite{volkov}
contains, however, the true potential, leading to an inconsistency 
of the values quoted in the Tables of \cite{volkov}.
\bibitem{footlim}
 When following the spirals towards their centers,
 we observe convergence of the mass, the charge and the frequency
 to finite limiting values. At the same time
 the functions approach limiting functions.
 These limiting functions consist of two parts.
 In the exterior part the limiting solutions 
 correspond to Schwarzschild solutions, 
 in the interior part we have not succeeded in determining
 the limiting solutions numerically.
\bibitem{footzero}
Trying to again use a scaling argument to construct the solutions
in the interval $[\omega_0(0), \omega_{\rm min}]$
in the limit $\kappa \rightarrow 0$, one can use $\hat r = \sqrt{\kappa} r$
to obtain a new set of two differential equations for the metric functions
together with a constraint for the scalar field, valid in an
interval $[0,\hat r_0]$, where $\hat r_0$ depends on the
frequency $\omega_s$.
Unfortunately, we have not succeeded in determining the dependence
$\hat r_0(\omega_s)$.
\end{thebibliography}
\end{document}